\documentclass[authoryear,preprint,review,12pt,5p,twocolumn]{elsarticle}
\usepackage{amssymb}
\usepackage{amsmath}
\usepackage[breaklinks]{hyperref}

\begin{document}

\begin{frontmatter}
\title{Dark matter admixed neutron stars with a realistic nuclear equation of state from chiral nuclear interactions}

\author[DipFis]{Domenico Scordino}
\author[DipFis,INFN]{Ignazio Bombaci}

\affiliation[DipFis]{organization={Dipartimento di Fisica ``E. Fermi'', Università di Pisa},
            addressline={Largo B. Pontecorvo 3}, 
            city={Pisa},
            postcode={56127}, 
            state={Italy},
            }
\affiliation[INFN]{organization={INFN, Sezione di Pisa},
            addressline={Largo B. Pontecorvo 3}, 
            city={Pisa},
            postcode={56127}, 
            state={Italy},
            }

\begin{abstract} 
We study the effects of dark matter on the structural properties of neutron stars. In particular we investigate how the presence of a dark matter component influences the mass-radius relation, the value of the maximum mass of a neutron star and other stellar properties.  
To model ordinary matter we use a state-of-the-art equation of state of $\beta$-stable nuclear matter obtained using the Brueckner-Hartree-Fock quantum many-body approach starting from two-body and three-body nuclear interactions derived from chiral effective field theory. 
The dark matter component of the star is modeled as a non-self-annihilating system of spin $1/2$ fermions at zero temperature and its equation of state as an ideal relativistic Fermi gas. The equilibrium configurations of these dark matter admixed neutron stars (DANS) are calculated by solving a generalization of 
the Tolman-Oppenheimer-Volkoff equations to the case where the system consists of two perfect fluids interacting solely through gravity. 
We find that, depending on the dark matter particle mass $m_\chi$, one can have somehow opposite effects on the stellar properties. 
In the case $m_\chi = 1\, \mathrm{GeV}$, the stellar gravitational maximum mass $M_{max}$ decreases, whereas in the case $m_\chi = 0.1\, \mathrm{GeV}$, $M_{max}$ increases with respect to the maximum mass of ordinary neutron stars. 
We also show that the presence of dark matter has indirect sizable effect on the proton fraction in the ordinary matter fluid and, in the case $m_\chi = 1\, \mathrm{GeV}$, results in a decrease of the threshold gravitational mass $M_{tot}^{durca}$ for having direct URCA processes and fast stellar cooling. 
Finally we study the stability of dark matter admixed neutron stars with respect to radial perturbations. 
\end{abstract}



\end{frontmatter}


\section{Introduction} 
\label{sec:intro}
In spite of its many successes, Einstein's theory of general relativity provides predictions that are at variance with astrophysical observational data on the kinematics of self-gravitating systems as galaxies and clusters of galaxies \citep{corbelli2000extended, nesti2023quest}. In order to reproduce these data using general relativity requires the introduction of a new type of matter, known as dark matter (DM), that is not directly observable except through its gravitational effects on ordinary matter. 
Cosmological data, currently explained by the $\Lambda$-cold dark matter ($\Lambda$CDM) model \citep{Peebels1982}, indicate that DM should be responsible for about 23\%  percent of the total energy density of the Universe while ordinary matter (OM) contributes only 4\%. The remaining portion is due to dark energy responsible for the accelerated expansion of the Universe \citep{GP1998,Garnavich_1998}.  

The main feature of DM is that its cross section for electromagnetic processes must be extremely small or zero. As a result, DM cannot lose energy by emitting electromagnetic radiation, as in the case of ordinary matter, and thus DM does not collapse at the center of galaxies. In fact, the rotation curve profiles of disc galaxies (i.e. the plot of the orbital speeds of visible stars or gas in the galaxy versus their radial distance from the galaxy's center) are in agreement with the presence of an extended and diffuse halo of DM (see \citep{sofue2013rotation, pato2015dynamical, lin2019dark}). 

None of the particles of the standard model meet the requirements for explaining the numerous astrophysical observations and cosmological data that require the existence of dark matter. Thus candidate particles to constitute dark matter most often emerge from extensions of the standard model. 

A much-studied class of such particles are WIMPs (weakly interacting massive particles), and many active experiments are attempting to reveal their existence. 
No such experiments, however, have so far obtained convincing results on the existence of DM particles even though very stringent constraints on the interaction cross section of dark matter with ordinary matter have been produced (see e.g. \citep{aprile2017first}). \\
If such a cross section is nonzero (although very small), it is conceivable that DM could be captured by stars or compact stellar objects, especially in environments such as the Galactic center, in a sufficient amount to produce observable effects. \\
Due to the very high densities ($10^{14}$ -- $10^{15}\ \mathrm{g/cm^3}$) in their cores, the extreme surface gravity ($g \sim 10^{12}\, \mathrm{m/s^2}$)  and their ages (up to $10^{10}\, \mathrm{yr}$), neutron stars are very good candidates where dark matter can be captured and can accumulate in sufficient amounts to influence their structural and evolutionary properties. Neutron stars can thus be used as natural detectors to identify DM and possibly unveil its nature
\citep{bertone2008compact, de2010neutron, kouvaris2010can,  kouvaris2011constraining, leung2011dark, Ciarcelluti2011, kouvaris2013dark, kouvaris2015asymmetric, tolos2015dark, ellis2018dark, deliyergiyev2019dark, ivanytskyi2020neutron, del2020change, kain2020radial, kain2021dark, Dengler:2021qcq, Shakeri-Karkevandi2024, Barbat:2024yvi, mariani2024constraining, thakur2024feasibility}. \\
In this paper we study the effects of the presence of DM on the structure and stability of neutron stars. In particular, we investigate, in a very detailed and systematic way, how the properties of these Dark matter Admixed Neutron Stars (DANS) are modified with respect to those of ordinary neutron stars in terms of the peculiarities and the fraction of DM present in the star. \\
The comparison of the results for the properties of DANS obtained in this work with some observational properties of neutron stars, such as the values of their gravitational mass (especially those of "massive" stars with $M \sim 2\, \mathrm{M_\odot}$ that place stringent constraints on the value of the 
Oppenheimer-Volkoff mass limit) or the combined mass and radius measurements (obtained from NASA's NICER X-ray observatory) allowed us to derive constraints on the mass of the dark matter particles and the fraction of dark matter present in the star. \\
This work is organized as follows. In Section~\ref{sec:struct}, we introduce the two-fluids Tolman-Oppenheimer-Volkoff (TOV) stellar structure equations. 
In Section~\ref{sec:eos}, we discuss the equation of state (EOS) for OM and for DM used in the present work. In Section~\ref{sec:DANS properties} we present results obtained by solving the two-fluids TOV equation and discuss the properties of DANS. In Section~\ref{sec:constraints on DM} we use observational data for properties of neutron stars to determine an upper limit to the dark matter particle mass. In Section~\ref{sec:stab}, we study the stability of DANS equilibrium configurations  with respect to small radial perturbations.
Finally, in Section~\ref{sec:conclusions} we give a brief summary of our work.  

\section{Structure equations for DANS}
\label{sec:struct}

Since the non-gravitational interaction between dark matter and ordinary matter is extremely small (e.g. \citep{bertone2005particle,aprile2017first}), it is possible to split the total energy-momentum tensor as the sum of the energy-momentum tensor of each of the two fluids (OM and DM) and to have covariant conservation for both of them.  
Accordingly, the equation of state of OM is independent on the state variables of DM and vice versa. In addition, it is assumed that each of the two fluids be a perfect fluid. Based on these assumptions, and further assuming a spherically symmetric and stationary distribution of OM and DM, the stellar structure equations in general relativity for DANS take the following form  (see e.g. \citep{kain2020radial}), which generalizes the TOV equations to the case of two fluids interacting exclusively through gravity:

\begin{align}
\label{2fTOV_a}
\frac{dP_j}{dr}  &=  -G \, \frac{m_{tot}(r)\, \varepsilon_j(r)}{c^2 r^2}\, \Bigg( 1 + \frac{P_j(r)}{\varepsilon_j(r)}\Bigg) \nonumber\\
           &\times \Bigg( 1 + \frac{4\pi r^3 P_{tot}(r)}{c^2 m_{tot}(r)}\Bigg)\, \Bigg( 1 - \frac{2\, G\, m_{tot}(r)}{c^2 r}\Bigg)^{-1}
\end{align}
and 
\begin{equation}
     \frac{d m_j(r)}{d r} = \frac{4 \pi}{c^2} r^2 \varepsilon_j(r) \,, 
\label{2fTOV_b}
\end{equation}  
where $G$ is the gravitational constant, $P_j$ and $\varepsilon_j$  (with $j =\, $OM,\,DM) are the pressure the and energy density for the OM and DM fluid, 
$m_j(r)$ is the gravitational mass enclosed within a sphere of radial coordinate $r$ (surface area $4\pi r^2$) for each of the two fluids, $m_{tot}(r) = m_{OM}(r) + m_{DM}(r)$ is the total gravitational mass enclosed within a sphere of radial coordinate $r$ and $P_{tot}(r) = P_{OM}(r) + P_{DM}(r)$ the total pressure.  

\subsection{Numerical integration}
To solve the stellar structure equations \eqref{2fTOV_a} and \eqref{2fTOV_b} we need to specify the equation of state for the two fluids (see next section) and 
the appropriate boundary conditions at the center ($r=0$) and at the surface ($r=R_j$, $j=OM, DM$) of the matter distribution for each fluid\,\citep{kain2020radial}:

\begin{equation*}
    m_j(0) = 0\,  \qquad   \varepsilon_j(0) = \varepsilon_{c,j} \,             
\end{equation*}
We define the radius $R_j$ of the distribution of fluid $j$ by the following condition
\begin{equation*}
    P_j(R_j) = P_j^{surf}
\end{equation*}
where $P_j^{surf}$ is a fixed value for the surface pressure of fluid $j$. For dark matter we use $P_{DM}^{surf}=0$ while for ordinary matter we chose 
$P_{OM}^{surf} = P_{OM}(\rho^*)$, where $\rho^* = 7.86\ \mathrm{g/cm^3}$ is the mass density of solid $^{56}\mathrm{Fe}$. \\
For $r> R_j$ we define $P_j(r) = 0$. The total radius of the star is 
\begin{equation}
    R = \max \{R_{OM}, R_{DM}\}
\end{equation}
Integrating Eq.(\ref{2fTOV_b}) we get the total gravitational mass $M_j$ for each of the two fluids ($j = OM, DM$)
\begin{equation}
      M_j \equiv m_j(R_j) =  \frac{4 \pi}{c^2} \int_{0}^{R_j} r^2 \varepsilon_j(r) \, dr 
\label{Mgrav_i}
\end{equation}  
and the total gravitational mass of the DANS is  
\begin{equation}
     M_{tot} = M_{OM} + M_{DM} \,.
\label{Mgrav_tot}     
\end{equation}

\section{Equation of state}
\label{sec:eos}
%
\subsection{The equation of state of ordinary matter}
\label{sec:BLeos}

Considering the existing uncertainties about the equation of state (EOS) of ordinary matter at high densities and considering that these uncertainties could masquerade the effects of dark matter on the observable properties of neutron stars, it is of paramount importance to employ an EOS for ordinary matter obtained using ab initio quantum many-body calculations well-grounded on modern theories of nuclear interactions. 

In this work we model the OM fluid of a DANS as a uniform electric-charge-neutral fluid of neutrons, protons, electrons, and muons in equilibrium with respect to the weak interaction ($\beta$-stable nuclear matter) at zero temperature ($T = 0$).

Recently a new microscopic EOS for this system has been obtained in Ref.~\citep{BL2018} (hereafter the BL EOS) for the zero temperature case, using the Brueckner-Hartree-Fock (BHF) quantum many-body approach (see \citep{BL2018} and references therein) starting from modern two-body and three-body nuclear interactions derived within chiral effective field theory (ChEFT) (e.g. \citep{2011PhR...503....1M, 2020RvMP...92b5004H}). 
These chiral nuclear interactions reproduce with high accuracy the nucleon-nucleon (NN) scattering data and the experimental binding energies of light (A = 3, 4) atomic nuclei \citep{LBK-2016}). 

This is a different approach with respect to other well-known microscopic EOS models for $\beta$-stable nuclear matter, as for example,  
the APR EOS \citep{Akmal:1998cf} or with respect to phenomenological EOS models derived, for example, from Skyrme nuclear interactions (see  \citep{Dutra2012} and references therein).

The BL EOS reproduces the empirical saturation point (i.e. saturation density $n_{0} = 0.16 \pm  0.01~{\rm fm}^{-3}$, and energy per nucleon 
$E/A|_{n_0} = -16.0 \pm 1.0~{\rm MeV}$) of symmetric nuclear matter, and other empirical properties (symmetry energy $E_{sym}$ and its slope parameter $L$, incompressibility) of nuclear matter at the saturation density $n_0$ (see Tab. 2 in \citep{BL2018}). \\
In particular the calculated $E_{sym}(n_0)$ and $L$ for this EOS fulfill the bounds imposed by the unitary Fermi gas limit \citep{Tews2017, Kievsky2018}. 
As has been shown in ref.~\citep{Tews2017} several EOS models currently used in neutron star structure calculations, in supernova explosions and binary neutron star mergers simulations violate the unitary Fermi gas bounds. 
The calculated $E_{sym}(n_0)$ and $L$ are also in agreement with the values $E_{sym}(n_0) = (38.1 \pm 4.7)\, \mathrm{MeV}$, 
$L = (106 \pm 37)\, \mathrm{MeV}$ extracted in Ref.~\citep{PhysRevLett.126.172503} from the measured neutron skin thickness of the $^{208}$Pb nucleus in the PREX-2 experiment \citep{PREX2}.

At supranuclear densities ($n > n_0$) the BL EOS model predicts a symmetry energy $E_{sym}(n)$ which is in very good agreement 
(see Fig. 3 in \citep{LBK-2016}) with the empirical constraints derived in \citep{danielewicz2014symmetry} using the excitation energies of 
isobaric analog states in nuclei.   
As is well known, the symmetry energy $E_{sym}(n)$, and particularly its density dependence is a crucial ingredient to determine the proton fraction in $\beta$-stable nuclear matter \citep{BL_1991, BaoAnLi_1,BaoAnLi_2,Bombaci_book_2022} and ultimately it has an impact on the value of stellar radius and on the thickness of the crust \citep{Lattimer-Prakash_2001, Steiner_2006} and on the possibility of having direct URCA processes \citep{durca1991} (see Section~\ref{sec:Core composition 1.4Msun}) and thus rapid cooling of neutron stars \citep{Page2004}. 
Further the BL EOS is consistent (see figure\ 2 in \citep{BL2018}) with the measured elliptic flow of matter in heavy-ion collisions experiments \citep{Danielewicz:2002pu}.

When computing static ordinary neutron star configurations the BL EOS (for the $\beta$-stable case) gives \citep{BL2018} a maximum mass $\rm{M}_{\rm max} = 2.08\, \mathrm{M_\odot}$, with a corresponding central density $\rho_c = 2.74 \times 10^{15}\, \mathrm{g/cm^3}$ and radius $\rm{R}(M_{\rm max}) = 10.26\, \mathrm{km}$  and a quadrupolar tidal polarizability coefficient $\Lambda_{1.4} = 385$ (for the $1.4\, \mathrm{M_\odot}$ neutron star \citep{Logoteta:2019cyb}) compatible with the constraints derived from GW170817 \citep{TheLIGOScientific:2017qsa}.
Recently, the BL EOS has been extended~\citep{Logoteta:2020yxf} to finite temperature and to arbitrary proton fractions. This finite-temperature EOS model has been applied to numerical simulations of binary neutron star mergers \citep{Bernuzzi:2020txg, Endrizzi:2018uwl, A.Prakash_2021}.  
Finally, to model the (ordinary matter) neutron star crust (i.e. for nucleonic density $\le 0.08$ fm$^{-3}$) we have used the Baym--Pethick--Sutherland \citep{bps71} and the  Negele--Vautherin \citep{NV73} EOS. 
%
\subsection{The equation of state of dark matter}
In the present work we consider non-self-annihilating fermionic DM. This so called fermionic asymmetric dark matter (ADM) \citep{KLZ_2009, Zurek_2014} carries 
a conserved charge which is analogous to the baryon number in the case of ordinary matter. We describe the ADM fluid as a non-interacting (ideal) gas of fermions with mass $m_\chi$ and spin $1/2$ \textbf{at $T = 0$}. 
The corresponding EOS is well known (see e.g. \citep{Shapiro-Teukolsky_1983}) and is given by the following expressions

\begin{equation*}
    \label{eos fermioni ideali}
    \begin{split}
        \varepsilon(x) &= \alpha(m_\chi, \nu)\chi(x) \\
        p(x) &= \alpha(m_\chi, \nu)\phi(x) \\
        n(x) &= n_0(m_\chi, \nu)x^3
    \end{split}
\end{equation*}

where $x\equiv \hbar k_F/(m_\chi c)$ is the adimensional Fermi momentum, the degeneracy factor $\nu = 2$ for spin $1/2$ particles and  

\begin{equation*}
    \label{equazione di stato sistema di fermioni ideali}
    \begin{split}
        &\alpha(m_\chi, \nu) = \nu \frac{\hbar c}{16\pi^2}\Bigl(\frac{m_\chi c^2}{\hbar c}\Bigr)^4 \\
        &n_0(m_\chi, \nu) = \nu \frac{(m_\chi c^2)^3}{6\pi^2(\hbar c)^3} \\
        &\chi(x) = x(1+x^2)^{1/2}(1+2x^2)-\ln[x+(1+x^2)^{1/2}] \\
        &\phi(x) = x(1+x^2)^{1/2}\Bigl(\frac{2}{3}x^2-1\Bigr)+\ln[x+(1+x^2)^{1/2}]
    \end{split}
\end{equation*}

\section{DANS properties}
\label{sec:DANS properties}

For a fixed values of the dark matter particle mass $m_\chi$, a solution of the two-fluid TOV equations \eqref{2fTOV_a} and \eqref{2fTOV_b} is uniquely 
identified once we fix the central mass densities (central energy densities) $\rho_{cOM}$ and $\rho_{cDM}$ for the two fluids.

\begin{figure}
    \centering
    \includegraphics[width=0.8\linewidth]{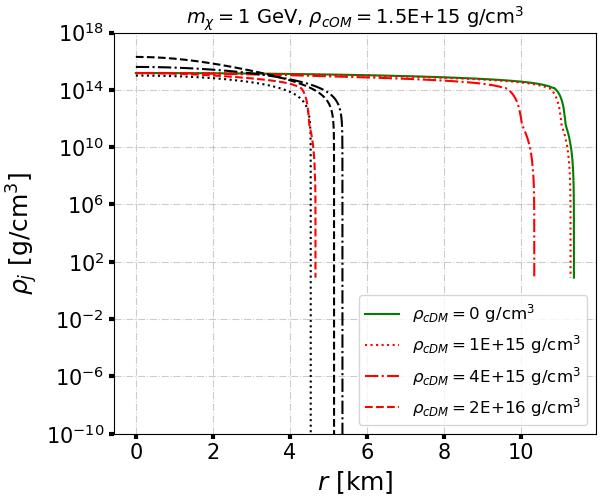}
    \caption{Mass-density profiles of ordinary matter (red) and dark matter (black) to the two-fluids TOV equation for $m_\chi = 1\ \mathrm{GeV}$, $\rho_{cOM} = 1.5\times 10^{15}\ \mathrm{g/cm^3}$ and different values of the dark matter central densities $\rho_{cDM}$. Each line-style correspond to a different choice of $\rho_{cDM}$: $(\rho_{cOM}, 0)$ (green continuous line), $(\rho_{cOM}, 10^{15}\ \mathrm{g/cm^3})$ (dotted lines), $(\rho_{cOM}, 4\times 10^{15}\ \mathrm{g/cm^3})$ (dash-dot lines), $(\rho_{cOM}, 2\times 10^{16}\ \mathrm{g/cm^3})$ (dashed lines). }
    \label{fig: Confronto profili rho 1GeV}
\end{figure}

\begin{table}[htp]
    \centering
    \begin{tabular}{c|ccccc}
        \hline\hline
        $\rho_{cDM}$ & $M_{OM}$ & $M_{DM}$ &  $f_{\chi}$ & $R_{OM}$ & $R_{DM}$\\
        $[\mathrm{g/cm^3}]$ & $[\mathrm{M_{\odot}}]$ & $[\mathrm{M_{\odot}}]$ & $[\%]$ & $[\mathrm{km}]$ & $[\mathrm{km}]$\\
        \hline
        0 & 1.895 & 0 & 0 & 11.375 & 0\\
        $10^{15}$ & 1.740 & 0.050  & 2.78 & 11.287 & 5.542 \\
        $4\times 10^{15}$ & 1.059 & 0.240 & 18.44 & 10.345 & 5.365\\
        $2\times 10^{16}$ & 0.095 & 0.453 & 82.62 & 4.664 & 5.148\\
        \hline\hline
    \end{tabular}
    \caption{Values for the gravitational mass of ordinary matter and dark matter and total fraction of dark matter obtained for $m_\chi = 1\ \mathrm{GeV}$, $\rho_{cOM} = 1.5\times 10^{15}\ \mathrm{g/cm^3}$ and different values of the dark matter central mass-density.}
    \label{tab:Confronto profili rho 1GeV}
\end{table}

\begin{figure*}
    \centering
    \includegraphics[width=0.95\linewidth]{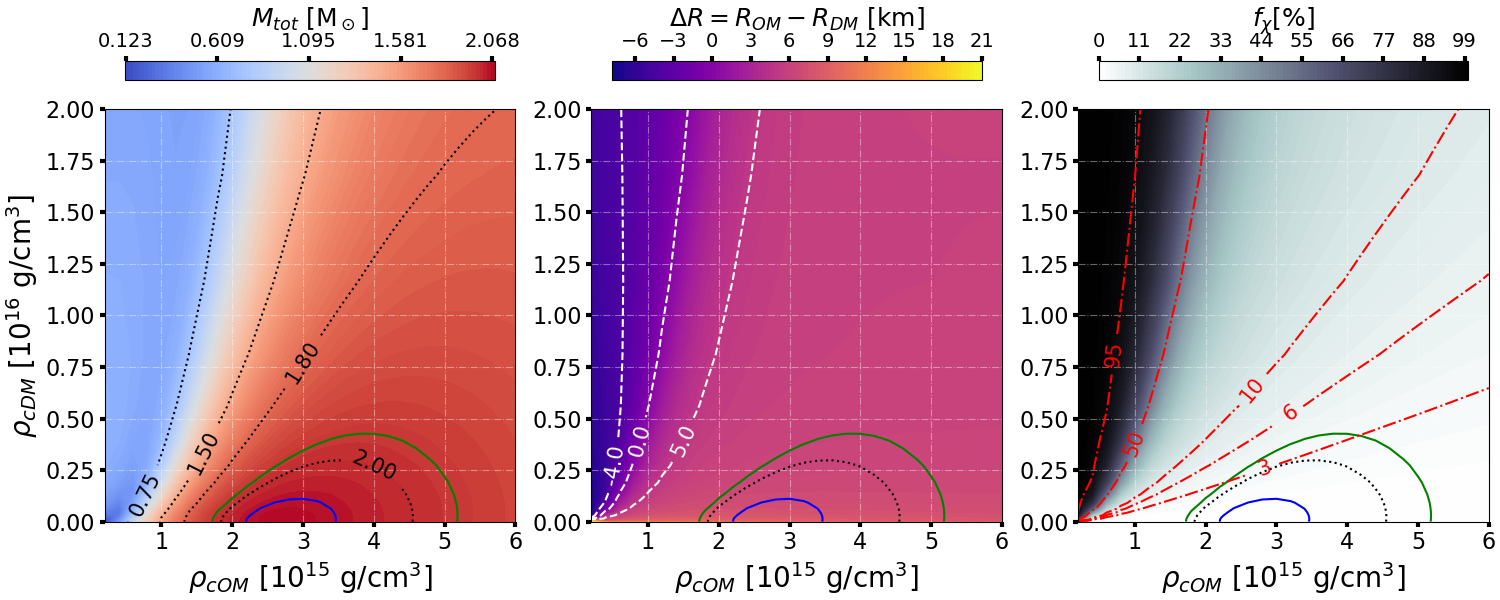}  
    \caption{Results for equilibrium configurations of DANS in the $(\rho_{cOM}, \rho_{cDM})$ plane for the case $m_\chi = 1\ \mathrm{GeV}$.   
    The left panel shows the total gravitational mass of the star $M_{tot} = M_{tot}(\rho_{cOM}, \rho_{cDM})$ for different choices of the central mass densities. 
    The central and right panel show the quantity $\Delta R = R_{OM} - R_{DM}$ and the stellar DM fraction in percent $f_\chi\,[\%]$ respectively as function of the central mass densities. 
    On each of the three panels we report the contour lines for $M_{tot}$ (left), $\Delta R$  (central) and $f_\chi\,[\%]$ (right). The green and blue lines in each panel delimit the region in the central mass density plane where the total gravitational mass $M_{tot}$ is compatible with the measurement of the mass of the neutron star associated with the pulsar PSR J0348+0432  ($2.01\pm 0.04\ \mathrm{M_\odot}$) \citep{antoniadis2013massive}.}     
    \label{fig: TOV2f plane-integration 1GeV}
\end{figure*}

\begin{figure*}
    \centering
    \includegraphics[width=0.95\linewidth]{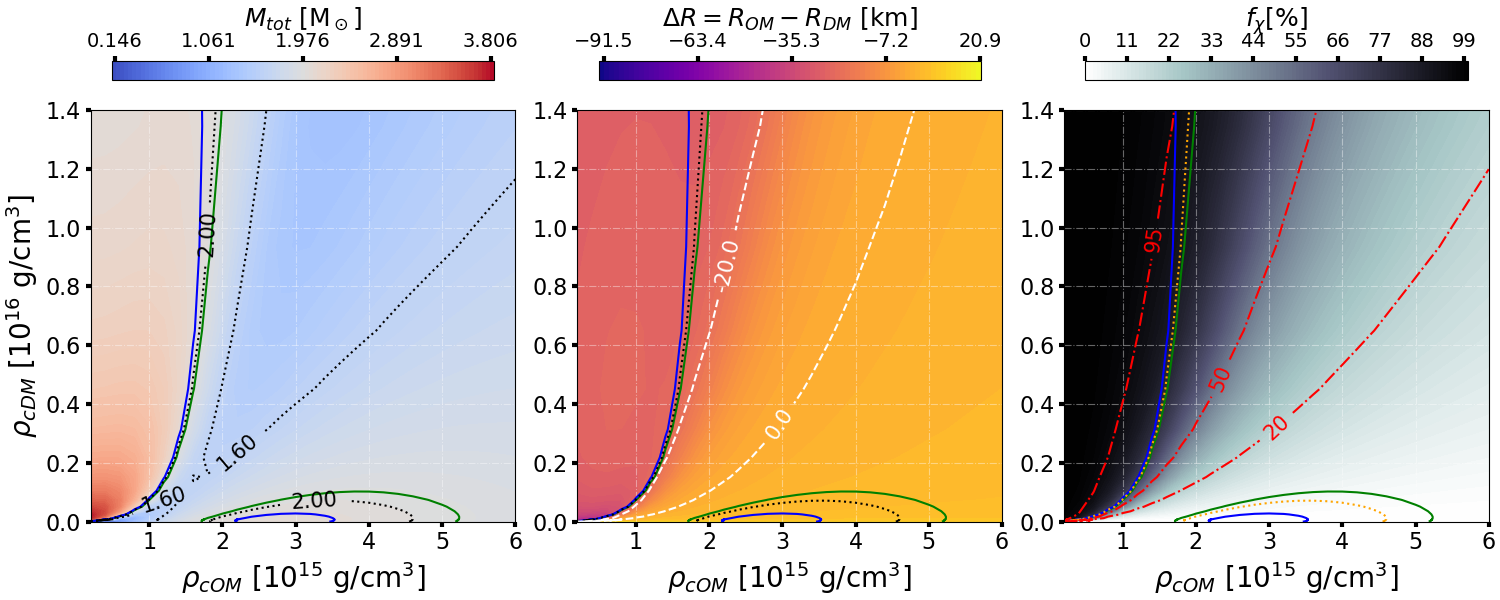}
  \caption{Same as in Fig \eqref{fig: TOV2f plane-integration 1GeV} but now for the case $m_\chi = 0.4\ \mathrm{GeV}$. }   
    \label{fig: TOV2f plane-integration 0.4GeV}
\end{figure*}

As an illustrative example, we display in Figure \eqref{fig: Confronto profili rho 1GeV} the radial mass density profiles $\rho_{OM}(r)$ for OM (red curves) and  $\rho_{DM}(r)$ for DM (black curves) obtained for a fixed value $\rho_{cOM}=1.5\times 10^{15}\ \mathrm{g/cm^3}$ of the OM fluid central density and using different values $\rho_{cDM}$ of the DM fluid central density. Each line-style in Figure \eqref{fig: Confronto profili rho 1GeV} corresponds to the profiles obtained for a different value of $\rho_{cDM}$ (see the figure caption for more details). As a baseline, we also plot in the same figure the mass density profile of an ordinary neutron star (green line). Taking the same stellar configurations considered in Figure \eqref{fig: Confronto profili rho 1GeV}, we report in 
Table \eqref{tab:Confronto profili rho 1GeV} the total gravitational masses (eq.\eqref{Mgrav_i}) of the OM and DM stellar components, the corresponding radii $R_{OM}$ and $R_{DM}$ and the stellar DM fraction $f_\chi$ defined as 

\begin{equation}
    \label{DMfrac}  
    f_\chi = \frac{M_{DM}}{M_{tot}}  
\end{equation}

All the results reported in Figure \eqref{fig: Confronto profili rho 1GeV} and in Table \eqref{tab:Confronto profili rho 1GeV} are relative to the case $m_\chi = 1\ \mathrm{GeV}$. 
Coming back to Figure \eqref{fig: Confronto profili rho 1GeV} we see that in the first two cases (dotted line and dash-dot lines), the radius of the dark matter distribution is smaller than the radius of the ordinary matter distribution ($R_{OM}\ge R_{DM}$) and thus we have a DANS with a dark matter core. In the third case reported in figure \eqref{fig: Confronto profili rho 1GeV} (dashed lines) we have $R_{OM}<R_{DM}$ and thus we have a DANS with a dark matter halo.

The results of our systematic calculations for hydrostatic equilibrium configurations of DANS in the $(\rho_{cOM}, \rho_{cDM})$ plane are reported in 
Fig. \eqref{fig: TOV2f plane-integration 1GeV} for the case $m_\chi = 1\ \mathrm{GeV}$. 
In particular, in the left panel we report, using a color scale, the total gravitational mass $M_{tot}$ of the star. The black dotted lines represent the contour lines ($M_{tot} = \mathrm{const}$), whereas the green and blue lines (reported in each of the three panels) are the contour lines which mark the region in the central mass densities plane where the calculated $M_{tot}$ is compatible with the measured mass ($2.01\pm 0.04\ \mathrm{M_\odot}$) \citep{antoniadis2013massive} of the neutron star associated to the pulsar PSR J0348+0432.  
In the central panel of Fig \eqref{fig: TOV2f plane-integration 1GeV}, we report, using a color scale, the difference $\Delta R = R_{OM} - R_{DM}$ between the radii of the OM and DM fluid distributions. The white dashed line labeled by "0.0" represents the contour line $\Delta R = 0$ and thus marks the boundary between DANS with a DM core (on the right of the $\Delta R = 0$ line) and those with a DM halo (on the left of the $\Delta R = 0$ line). Finally, in the right panel of Fig \eqref{fig: TOV2f plane-integration 1GeV} we plot the DM fraction (in percent) $f_\chi\,[\%]$ and the contour lines for this quantity (red dash-dotted lines). The results in the right panel of Fig \eqref{fig: TOV2f plane-integration 1GeV} ($m_\chi = 1\, \mathrm{GeV}$) clearly show that only DANS with a low DM fraction ($f_{\chi} < 4\, \%$) are compatible with the measured mass of the neutron star associated with PSR J0348+0432 and that these configurations are all characterized by the presence of a DM core (see middle panel). 

In Fig. \eqref{fig: TOV2f plane-integration 0.4GeV} we show our results for DANS properties in the case $m_\chi = 0.4\ \mathrm{GeV}$. Now, in addition to the region of the $(\rho_{cOM}, \rho_{cDM})$ plane populated by DANS with a DM core, with a low DM fraction ($f_\chi \lesssim 3.3\,\%$) and having $M_{tot} \sim 2\, \mathrm{M_\odot}$, there is a second region of the $(\rho_{cOM}, \rho_{cDM})$ plane populated by DANS with $M_{tot} \gtrsim 2\, \mathrm{M_\odot}$. All members of this second group of DANS have an extended DM halo and are almost completely formed by DM. 
        
The possibility of having DANS with a large DM fraction or compact stars made entirely by fermionic DM and the possible mechanisms to form them have been discussed in several studies (e.g. \citep{narain2006compact, kouvaris2015asymmetric, gresham2019asymmetric}). 


As a general trend, we have that for fixed values of the central mass densities of the two fluids, higher values of $m_\chi$ tend to favor the formation of a dark matter core, whereas lower values of $m_\chi$ tend to favor the formation of a dark matter halo.

\subsection{Mass-radius relation for fixed DM fraction}
\label{sec:M-R relation}
To investigate the effects of DM on observable properties of DANS, we consider stellar sequences with a fixed DM fraction $f_{\chi}$. In other words, we solve the two-fluid structure equations by taking the central mass density of the OM fluid $\rho_{cOM}$ and  the DM fraction $f_{\chi}$ as independent variables. 

\begin{figure*}
    \centering
    \includegraphics[width=0.8\linewidth]{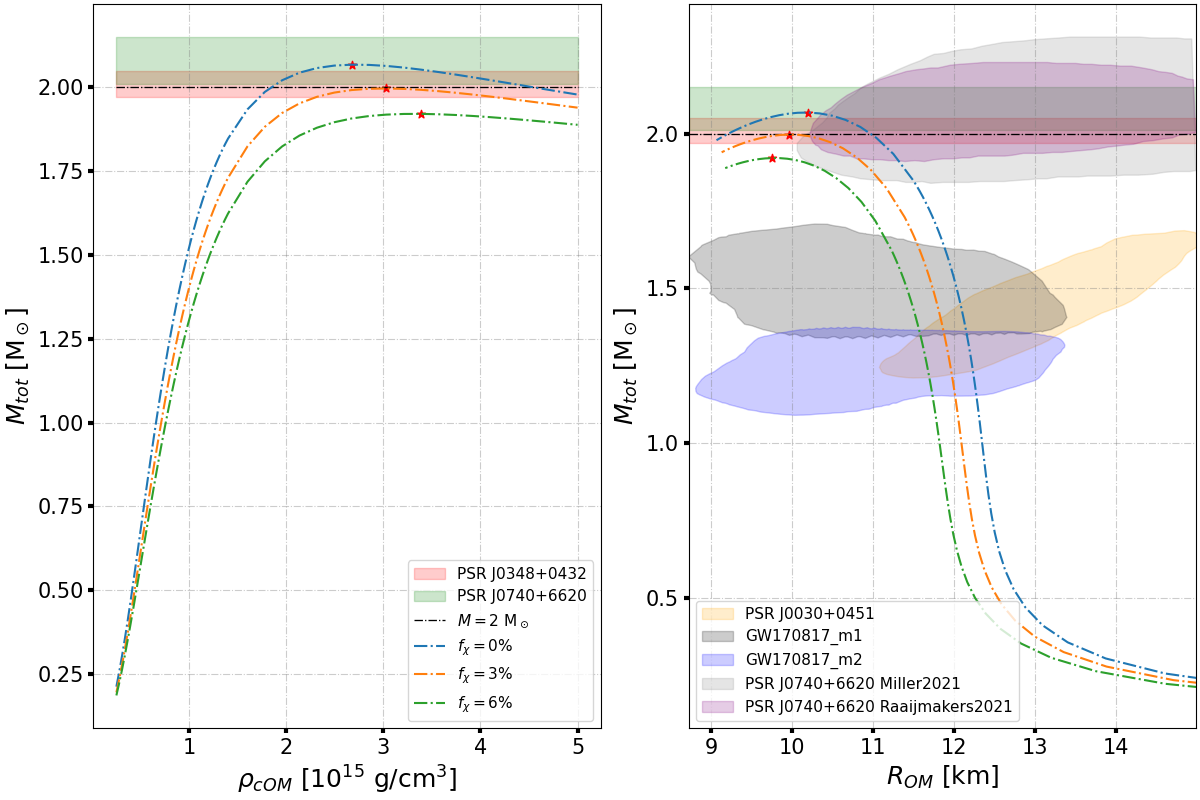}
    \caption{Total gravitational mass of the star $M_{tot}$ as a function of $\rho_{cOM}$ (left panel) and $M_{tot}$ as a function of the radius $R_{OM}$ of the OM fluid distribution (right panel) for different values of the DM fraction ($f_\chi = 0\,\%,  3\,\%, 6\,\%$) and in the case $m_\chi = 1\ \mathrm{GeV}$. 
    The red star symbol on the top of each curve marks the maximum mass configuration.  
    The light green and the pink horizontal bands represent respectively the measured mass of the neutron star associated to the pulsar PSR J0740+6620 \citep{Fonseca:2021wxt} and to PSR J0348+0432 \citep{antoniadis2013massive}. In the right panel, the orange region marks the NICER measurement for the mass and radius of the neutron star in PSR J0030+0451 \citep{miller2019psr} while the plum and light gray regions in the top-right part of the figure mark the NICER measurement for the mass and radius of the neutron star in PSR J0740+6620 obtained respectively in \citep{raaijmakers2021constraints} and in \citep{Miller:2021qha}. The gray and blue regions represent the portion of the mass-radius plane compatible with the data obtained from the GW170817 event (at $90\%\, CL$) \citep{GW170817_MR}.} 
    \label{fig: Plot MR fix fchi 1GeV}
\end{figure*}

In figure \eqref{fig: Plot MR fix fchi 1GeV}, for the case $m_\chi = 1\ \mathrm{GeV}$, we show the total gravitational mass of the star $M_{tot}$ as a function of $\rho_{cOM}$ (left panel) and $M_{tot}$ as a function of the radius $R_{OM}$ of the OM fluid distribution (right panel). The surface of the OM fluid distribution is in fact the place where the electromagnetic radiation, that allows to determine $R_{OM}$, is generated.

The light blue dash-dotted curve in both panels represents the ordinary neutron star sequence ($f_{\chi} = 0$). The other two curves represent DANS sequences with $f_{\chi} =  3 \%$ (orange dash-dotted lines) and $f_{\chi} =  6 \%$ (green dash-dotted lines). These two stellar sequences represent DANS having a DM core (see central and right panels in Fig \eqref{fig: TOV2f plane-integration 1GeV}). The light green horizontal band represents the measured mass $(2.08\pm 0.07)\ \mathrm{M_\odot}$ \citep{Fonseca:2021wxt} of the neutron star associated to the pulsar PSR J0740+6620, whereas the pink horizontal band represents the measured mass $(2.01\pm 0.04)\ \mathrm{M_\odot}$ \citep{antoniadis2013massive} of the neutron star associated to the pulsar PSR J0348+0432. 
As we can see from the results reported in figure \eqref{fig: Plot MR fix fchi 1GeV}, the presence of DM  with $m_\chi = 1\ \mathrm{GeV}$ reduces the value of the maximum gravitational mass $M_{max} \equiv M_{tot,max}$ of the star with respect to the one for ordinary ($f_\chi = 0$) neutron stars.
Considering DANS sequences with $f_\chi \gtrsim 3\%$ the calculated maximum mass is no longer compatible with the present measured masses of "heavy" neutron stars with $M \sim 2\, \mathrm{M_\odot}$ \citep{Fonseca:2021wxt,antoniadis2013massive}. We also note that the $M_{tot}(\rho_{cOM})$ curve is shifted towards higher values of the ordinary matter central density $\rho_{cOM}$ compared to the ordinary neutron stars case ($f_\chi = 0\%$) as $f_\chi$ increases.

The presence of DM also affects the value of the OM radius $R_{OM}$ (right panel) making the OM distribution more compact \citep{ellis2018dark, ivanytskyi2020neutron}. For example, considering an ordinary neutron star with a gravitational mass $M = 1.4\ \mathrm{M_\odot}$, a DM fraction $f_{\chi} =  6 \%$ reduces the stellar observable radius $R_{OM}$ by $\sim 0.6\ \mathrm{km}$  i.e by $\sim 5\, \%$. These changes in the stellar radius are within the bounds of uncertainty of present instruments, as NICER, for measuring neutron star radii. 
The orange region, in the right panel of Fig. \eqref{fig: Plot MR fix fchi 1GeV}, marks the NICER measurement for the mass and radius of the neutron star in PSR J0030+0451 \citep{miller2019psr}, while the plum and light gray regions in the top-right part of the figure mark the NICER measurement for the mass and radius of the neutron star in PSR J0740+6620 obtained respectively in \citep{raaijmakers2021constraints} and in \citep{Miller:2021qha}. The gray and blue regions represent the portion of the mass-radius plane compatible with the data obtained from the GW170817 event (at $90\%\, CL$) \citep{GW170817_MR}.

%

\begin{figure*}
    \centering
    \includegraphics[width=0.8\linewidth]{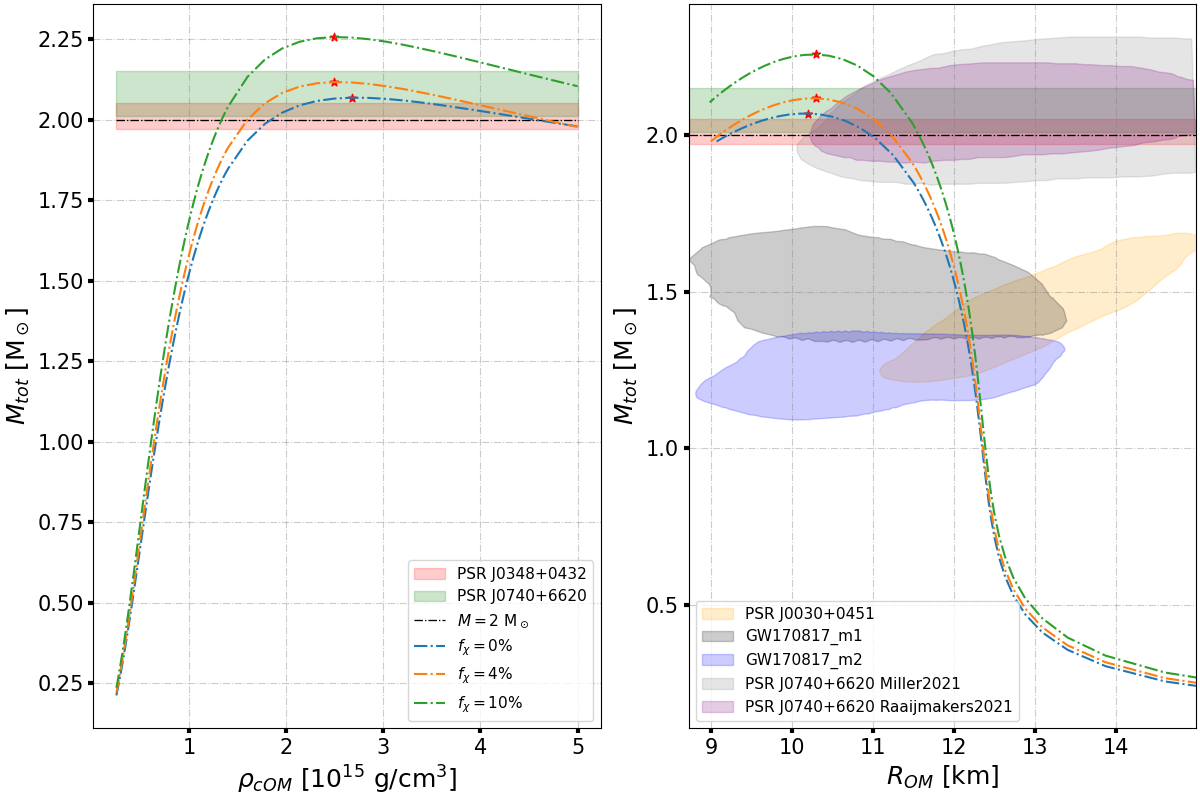}
    \caption{Same as in Fig \eqref{fig: Plot MR fix fchi 1GeV} but now for the case $m_\chi = 0.1\ \mathrm{GeV}$ and DM fraction $f_\chi = 0\,\%,  4\,\%, 10\,\%$.}
    \label{fig: Plot MR fix fchi 0.1GeV}
\end{figure*}

In Fig. \eqref{fig: Plot MR fix fchi 0.1GeV} we report DANS sequences for DM particles having a mass $m_\chi = 0.1\ \mathrm{GeV}$ and for fixed DM fraction. 
In this case, we get stars with a DM halo. 
This leads to a growth of the total gravitational mass of the star as $f_\chi$ increases for every considered value of the central density of the ordinary matter distribution and, in particular, to a growth of the maximum gravitational mass of the $M(R)$ sequence.


\begin{figure}
    \centering
    \includegraphics[width=0.8\linewidth]{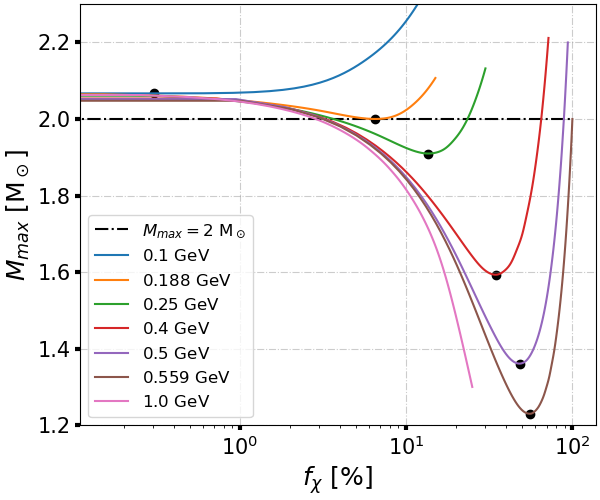}
    \caption{Maximum mass for dark matter admixed neutron stars as a function of the  dark matter fraction $f_\chi$ for fixed values of the dark matter particle mass $m_\chi$. The black heavy dot on top of each curve denotes the minimum of each curve.}
    \label{fig: Mmax_fD}
\end{figure}

To investigate in detail how the maximum mass $M_{max}$ of DANS depends on the DM fraction $f_\chi$ and on the DM particle mass $m_\chi$, we show in Fig. \eqref{fig: Mmax_fD} our results for the function $M_{max}(m_\chi, f_\chi)$ calculated for a few constant values of $m_\chi$ (curves in different colors). The black heavy dot on top of each curve denotes the minimum of the function $M_{max}(f_\chi)$ for fixed $m_\chi$. 
We see that for $m_\chi \approx 0.188\ \mathrm{GeV}$ the minimum of the orange curve is equal to $2\ \mathrm{M_\odot}$ (a value we take as a reference for the measured masses of 'heavy' neutron stars): this result means that for $m_\chi \le 0.188\ \mathrm{GeV}$ we obtain a value for $M_{max}$ that is always compatible with the measured masses of neutron stars for any value of $f_\chi$. 

For any fixed value of $m_\chi \ge 0.188\ \mathrm{GeV}$, we define a critical value of the DM fraction, and denote it with 
$f_\chi^{crit} = f_\chi^{crit}(m_\chi)$, as the solution of the equation $M_{max}(f_\chi, m_\chi) = 2\ \mathrm{M_\odot}$  \citep{ivanytskyi2020neutron}. 
From our results in Fig. \eqref{fig: Mmax_fD} we see that for $0.188\, \mathrm{GeV} < m_\chi \leq  0.559\, \mathrm{GeV}$ there are
two solutions that satisfy equation $M_{max}(f_\chi,m_\chi) = 2\, M_\odot$. 
Thus, for example, in the case $m_\chi = 0.4\ \mathrm{GeV}$ (red curve in our Fig. \eqref{fig: Mmax_fD}) there are two groups of DANS  which satisfy the condition $M_{max}(f_\chi,m_\chi) \geq 2\, M_\odot$, i.e. have masses in agreement with present measured neutron star masses.
The first of these two groups is represented by DANS with a low DM fraction ($f_\chi \leq 3.27\%$) and having a DM core, whereas the second group consist of DANS with a high DM fraction ($f_\chi \ge 64.93\%$) and having a DM halo. 
These results are in line with our findings in Fig.~\eqref{fig: TOV2f plane-integration 0.4GeV}. 
The value $m_\chi = 0.559\, \mathrm{GeV}$ marks the transition from two to one solution for the critical DM fraction $f_\chi^{crit}$. 
In fact, for this value of the DM particle mass the pure ($f_\chi = 100\%$) DM compact star sequence has a maximum mass $M_{max}(f_\chi=100\%) = 2.00\, M_\odot$ \citep{narain2006compact}. 
Our results are thus in line and extend those from \citep{ivanytskyi2020neutron}, where a different ordinary matter EOS was used.

%

\begin{figure}
    \centering
    \includegraphics[width=0.8\linewidth]{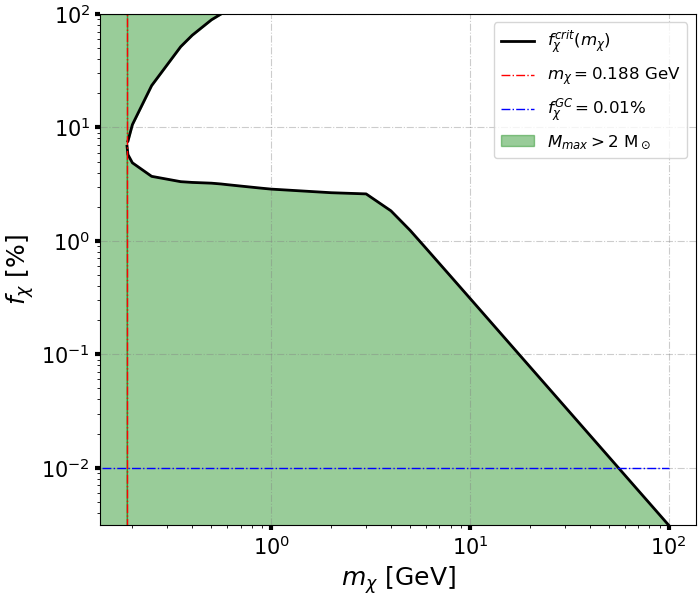}
    \caption{Critical value of $f_\chi$ as a function of $m_\chi$. In the green region and along the black line the maximum mass for DANS is compatible with observation. The vertical red dashed line represents the value $m_\chi = 0.188\, \mathrm{GeV}$. Below that value, the maximum gravitational mass obtained for the $M_{tot}(R_{OM})$ sequence is always compatible with the measured masses of neutron stars for any value of $f_\chi$ (see also Fig. \ref{fig: Mmax_fD}).
    The horizontal blue dashed line indicates a conservative estimate of the dark matter fraction that a $2\ \mathrm{M_\odot}$ neutron star in the Galactic central region can attain due to dark matter accretion during both the main sequence (MS) stellar phase and the neutron star phase 
    (see Sec.\ref{sec:constraints on DM}). 
    }
    \label{fig: m_fD_2Msun}
\end{figure}

The critical DM fraction is shown in Fig. \eqref{fig: m_fD_2Msun} by the heavy black curve labeled $f_\chi^{crit}(m_\chi)$. The green region in the same figure thus represents the portion of the $(m_\chi, f_\chi)$ plane where the calculated maximum mass for DANS is compatible with the present measured masses of "heavy" neutron stars with $M_{tot} \sim 2\, \mathrm{M_\odot}$ \citep{Fonseca:2021wxt,antoniadis2013massive}. It is useful to observe that, in the range  $10 \leq  m_\chi/ \mathrm{GeV} \leq 100$ the critical DM curve $f_\chi^{crit}(m_\chi)$ is well described by the following equation

\begin{equation}
    \label{fit}
    f_\chi^{crit}(m_\chi) = 31.15\Bigl(\frac{1\ \mathrm{GeV}}{m_\chi}\Bigr)^2+6.62\times 10^{-6}
\end{equation}

\subsection{Proton fraction and direct URCA processes in DANS}
\label{sec:Core composition 1.4Msun}

As we show in this subsection, the presence of DM influences not only the bulk properties (mass, radius, maximum mass) of a neutron star, but in addition has indirect sizable effect (through gravitational interaction) on the particle fractions of the OM fluid constituents when considering DANS with a fixed total gravitational mass. In particular the value of $f_\chi$ influences the proton fraction $x_p = n_p/n$ (with $n = n_n + n_p$  being the baryon number density) in $\beta$-stable nuclear matter. 

As has been shown in Ref. \citep{durca1991}, when the proton fraction $x_p$ is larger than a threshold value, denoted as $x_p^{durca}$, the so-called direct URCA processes $n \rightarrow p + e^- + \bar{\nu}_e\,$, $~ p + e^-  \rightarrow n +  \nu_e\,,$ can occur in neutron star matter. 
The direct URCA processes enhance neutrino emission and neutron star cooling rates by a very large factor compared to the so-called modified URCA processes $(n, p) + n \rightarrow p + e^- + \bar{\nu}_e\,$,   $~(n, p) + p + e^-  \rightarrow n +  \nu_e\,,$ causing a fast cooling of the star \citep{durca1991,Page2004}. 

In $\beta$-stable nuclear matter the threshold proton fraction for direct URCA processes can be written \citep{BL2018}  

\begin{equation}
    \label{durca}
    x_p^{durca} = \frac{1}{1 + \big(1 + Y_e^{1/3}\big)^3} \, ,
\end{equation} 

where $Y_e = n_e/(n_e + n_\mu)$ is the leptonic electron fraction. Notice that below the muon threshold density $x_p^{durca} = 1/9$ while for asymptotically large baryon densities $x_p^{durca} \sim 14.77\, \%$. 

\begin{figure}
    \centering
    \includegraphics[width=0.9\linewidth]{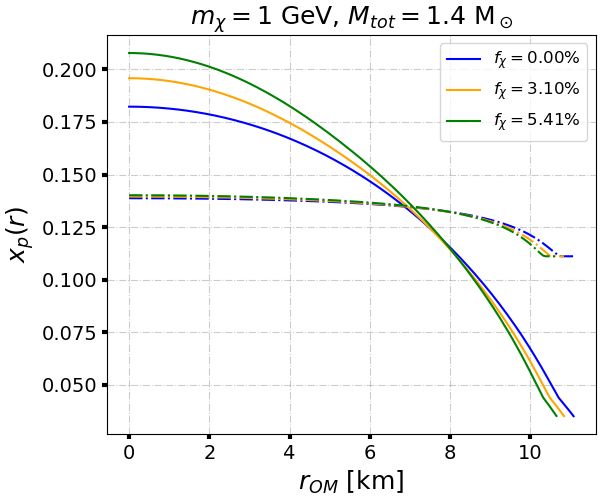}
    \caption{Proton fraction $x_p(r)$ (continuous lines) and direct URCA processes threshold (dash-dotted lines) $x_p^{durc}(r)$ as functions of the radial coordinate in the ordinary matter fluid for a DANS having a gravitational mass $M_{tot}=1.4\ \mathrm{M_\odot}$ in the case $m_\chi = 1\ \mathrm{GeV}$ and different values of $f_\chi$.}
    \label{fig: xp_durca 1.4Msun 1GeV}
\end{figure}

In Fig.\ \ref{fig: xp_durca 1.4Msun 1GeV}, we plot the proton fraction $x_p$ in $\beta$-stable nuclear matter as a function of the radial coordinate $r_{OM}$ in a DANS having a total gravitational mass $M_{tot} = 1.4\, \mathrm{M_\odot}$ and considering three different values, $f_\chi = 0\,\%,\ 3.1\,\%,\ 5.41\,\%$, for the DM fraction (continuous lines) and taking $m_\chi = 1\, \mathrm{GeV}$. Each of the three dash-dotted lines represents the threshold proton fraction $x_p^{durca}$ for the direct URCA processes associated to the $x_p$ curve with the same $f_\chi$ (curves with the same color). The intersection point between each couple of curves with the same color gives the value of the radial coordinate $r_{OM}^{durca}$ below which direct URCA processes are possible.  

As is clearly seen in Fig.\ \ref{fig: xp_durca 1.4Msun 1GeV}, the presence a DM core produces a significant increase in the proton fraction (particularly in the central region of the star) and small increase in the value of $r_{OM}^{durca}$ . 
Notice that in the case of an ordinary neutron star ($f_\chi = 0$) with mass $M = 1.4\ \mathrm{M_\odot}$ described by the BL EOS, direct URCA processes are already possible \citep{BL2018}. 
Complementary to Fig.\ \ref{fig: xp_durca 1.4Msun 1GeV}, we report in Tab.\ \ref{tab: durca2 1.4Msun} some of the properties of a DANS with $M_{tot}=1.4\ \mathrm{M_\odot}$ in the case $m_\chi = 1\ \mathrm{GeV}$. 

\begin{figure}
    \centering
    \includegraphics[width=0.9\linewidth]{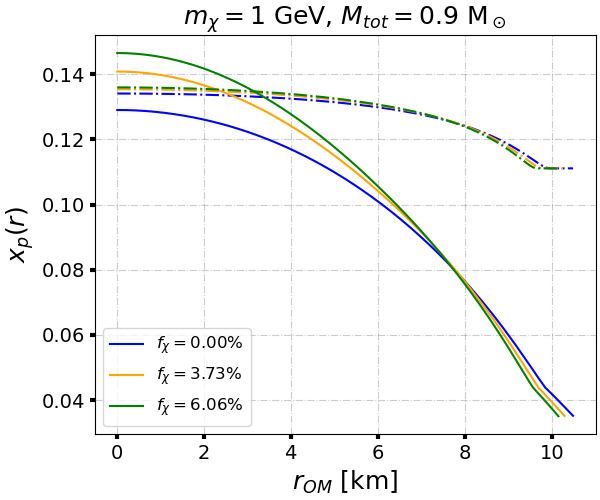}
    \caption{Same as in Fig.\ \ref{fig: xp_durca 1.4Msun 1GeV} but now for $M_{tot}=0.9\ \mathrm{M_\odot}$.}
    \label{fig: xp_durca 0.9Msun 1GeV}
\end{figure}

Fig.~\ref{fig: xp_durca 0.9Msun 1GeV} displays the same physical quantities as in the previous figure, but this time for DANS with a total gravitational mass $M_{tot}=0.9\ \mathrm{M_\odot}$. 
Now in the case of ordinary neutron stars direct URCA processes are not possible ($x_P < x_p^{durca}$), but they can be switched on if a sufficient amount of DM is present in the star.
Complementary to Fig.\ \ref{fig: xp_durca 0.9Msun 1GeV}, we report in Tab.\ \ref{tab: durca2 0.9Msun} some of the properties of a DANS with $M_{tot} = 0.9\ \mathrm{M_\odot}$ in the case $m_\chi = 1\ \mathrm{GeV}$. 

\begin{table}
    \centering
    \begin{tabular}{c|cccc}
        \hline\hline
        $f_\chi$ & $\rho_{cOM}$ & $n(r=0)$ & $x_p(r=0)$ & $r_{OM}^{durca}$ \\
        $[\%]$ & $[10^{15}\ \mathrm{g/cm^3}]$ & $[\mathrm{fm^{-3}}]$ & & $[\mathrm{km}]$ \\
        \hline
        0 & 0.910 & 0.497 & 0.182 & 6.84 \\
        3.10 & 1.000 & 0.539 & 0.196 & 6.94 \\
        5.41 & 1.086 & 0.578 & 0.208 & 7.08\\
        \hline\hline
    \end{tabular}
    \caption{Central mass and baryon number density (first and second column), central proton fraction (third column) and the radial coordinate $r_{OM}^{durca}$ below which direct URCA processes are possible in a $1.4\ \mathrm{M_\odot}$  DANS for different value of the dark matter fraction $f_\chi$ (first column). All the results are relative to the case $m_\chi = 1\ \mathrm{GeV}$.}
    \label{tab: durca2 1.4Msun}
\end{table}

\begin{table}
    \centering
    \begin{tabular}{c|cccc}
        \hline\hline
        $f_\chi$ & $\rho_{cOM}$ & $n(r=0)$ & $x_p(r=0)$ & $r_{OM}^{durca}$ \\
        $[\%]$ & $[10^{14}\ \mathrm{g/cm^3}]$ & $[\mathrm{fm^{-3}}]$ & & $[\mathrm{km}]$ \\
        \hline
        0 & 6.076 & 0.345 & 0.129 & / \\
        3.73 & 6.688 & 0.377 & 0.141 & 2.40 \\
        6.06 & 6.993 & 0.393 & 0.147 & 3.13 \\
        \hline\hline
    \end{tabular}
    \caption{Same as Tab.\ \ref{tab: durca2 1.4Msun} but for $M_{tot}=0.9\ \mathrm{M_\odot}$.}
    \label{tab: durca2 0.9Msun}
\end{table}

\begin{table}
    \centering
    \begin{tabular}{c|cc}
        \hline\hline
        $f_\chi\ [\%]$ & $M_{tot}^{durca}\ [\mathrm{M_\odot}]$ & $M_{max}\ [\mathrm{M_\odot}]$ \\
        \hline
        0 & 0.96 & 2.07 \\
        3.10 & 0.87 & 1.99 \\
        5.41 & 0.82 & 1.94 \\
        \hline\hline
    \end{tabular}
    \caption{Values for $M^{durca}_{tot} = M_{tot}(n^{durca})$ (second column) and for the maximum gravitational mass $M_{max}(f_\chi)$ 
    (second column) for DANS sequences with different values of dark matter fraction $f_\chi$. 
    The calculated values for the threshold baryon number density for having direct URCA processes is $n^{durca} = 0.361\ \mathrm{fm^{-3}}$.      
    All the results are relative to the case $m_\chi = 1\ \mathrm{GeV}$.} 
    \label{tab: Mdurca 1.4Msun}
\end{table}

The calculated values for the threshold baryon number density $n^{durca}$ for having direct URCA processes and the corresponding threshold proton fraction $x_p(n^{durca})$ do not depend on the DM fraction, since these two quantities are exclusively determined by the nuclear interactions and particularly are strongly affected by the density dependence of the nuclear symmetry energy \citep{BL_1991, Bombaci_book_2022}.  
In the case of the BL EOS one has $n^{durca} = 0.361\ \mathrm{fm^{-3}}$ and $x_p(n^{durca}) = 0.1347$ \citep{BL2018}.  
The stellar total gravitational mass $M_{tot}^{durca} \equiv M_{tot}(n_c = n^{durca})$ beyond which direct URCA processes are possible  depends instead  on the DM fraction. Its value together with the value of the gravitational maximum mass $M_{tot}^{max}$ is reported in Tab.\ \ref{tab: Mdurca 1.4Msun} for some values of the DM fraction. As we see the presence of a DM core reduces the value of $M_{tot}^{durca}$ and thus makes possible fast cooling of "light" neutron stars. 
Our results are in general agreement with what reported in the recent papers \citep{avila2024rapid,giangrandi2024impact}. 
For example, we find that a DM fraction $f_\chi \sim 3\%$ produces a decrease in the critical mass $M_{tot}^{durca}$ for having direct URCA process of about $0.09\,M_\odot$, while the authors of reference \citep{avila2024rapid,giangrandi2024impact} find a decrease between 0.10 and 0.13 $M_\odot$ for the two cases of ordinary matter EOS they used. 
However, we find a significant difference when we compare our calculated values for $M_{tot}^{durca}(f_\chi)$  
with those reported in \citep{avila2024rapid,giangrandi2024impact}. 
For example, we predict that a DM fraction $f_\chi = 3.1\%$ could produce rapid cooling of neutron stars with a quite low mass 
($M_{tot} \geq 0.87\,M_\odot$) whereas in \citep{avila2024rapid,giangrandi2024impact}, it is found  $M_{tot}^{durca} = 1.80\,M_\odot$ for the same DM fraction and in the case of the phenomenological Induced Surface Tension (IST) EOS model for ordinary matter. 

We are able to trace the origin of these substantial differences between our results and those of \citep{avila2024rapid,giangrandi2024impact}. 
In fact, large differences in the calculated  threshold stellar mass $M_{tot}^{durca}$ for having direct URCA processes are already present when we compare our results (see also Table 4 in \citep{BL2018}) with those of Ref. \citep{avila2024rapid,giangrandi2024impact} in the case of  ordinary neutron stars ($f_\chi = 0$). Definitely, these differences are due to the use of distinct EOS models for describing ordinary matter and, as we argue below, they are due to a different density dependence of the nuclear symmetry energy in the considered nuclear matter EOS models. 

As already discusses in Section~\ref{sec:BLeos}, the value of the proton fraction $x_p$ in $\beta$-stable nuclear matter (and the threshold baryon number density $n^{durca}$ for having direct URCA process) is highly related to the density dependence of the symmetry energy $E_{sym}(n)$.  
A stiff symmetry energy, as suggested by the PREX-2 measurement \citep{PREX2} of the neutron skin thickness  of the $^{208}$Pb nucleus, 
favors a large proton fraction and a low value ($M_{tot}^{durca} \lesssim 1.45\, M_\odot$) for the threshold stellar mass for direct URCA processes (see Fig. 3 in \citep{PhysRevLett.126.172503}), in agreement with our results for ordinary ($f_\chi = 0$) neutron stars. 

In the case of the IST EOS model, used in \citep{avila2024rapid,giangrandi2024impact}, the symmetry energy $E_{sym}(n)$ is softer than the one for the BL model and as a consequence the proton fraction $x_p(n)$ in $\beta$-stable matter in the case of the IST EOS is much lower than the one of the BL EOS 
(compare Fig.~4 in \citep{avila2024rapid} with Fig.~3 in \citep{BL2018}).  
Note, also, that for the case of the  IST and FSU2R EOS models, used in \citep{avila2024rapid,giangrandi2024impact}, the symmetry energy $E_{sym}(n_0)$ and its slope $L$, at nuclear saturation density, are in disagreement with the values derived by the measured neutron skin thickness of the $^{208}$Pb nucleus by the PREX-2 experiment. In addition these two physical quantities violate the bounds imposed by the unitary Fermi gas limit \citep{Tews2017} in the case of the IST EOS.  

Besides, in calculating the proton fraction in $\beta$-stable nuclear matter the authors of references 
\citep{avila2024rapid, giangrandi2024impact} do not include negative muons ($\mu^-$).   
In the BL EOS model muons appear at a baryon density $n = 0.123\ \mathrm{ fm^ {-3}}$ (i.e. below the nuclear matter saturation density). 
Any reliable calculation of the proton fraction in $\beta$-stable nuclear matter requires the presence of muons.  

Therefore, our results on the effect of dark matter on the proton fraction in DANS and the possibility of having fast cooling already 
in stars with a mass below one solar mass are very interesting and different from what has been reported in the literature so far.

\section{Constraints from astrophysical observations on DM properties}
\label{sec:constraints on DM}

As discussed in \citep{ivanytskyi2020neutron}, it is possible to give an upper limit to the mass $m_\chi$ of 
the dark matter particles by using the results reported in Fig.\,\eqref{fig: m_fD_2Msun} and comparing them with the estimate of the amount of dark matter that can be accumulated onto a neutron star. 
To make such an estimate, it is necessary to evaluate the dark matter accretion rate at each evolutionary stage of the star, from the formation of the progenitor star up to the stable neutron star phase. 

The accretion of DM on massive stars which form neutron stars, has been most carefully investigated during their main sequence phase \citep{scott2009dark, lopes2019asteroseismology} and during the stable neutron star phase \citep{kouvaris2008wimp,kouvaris2011constraining}.

Dark matter accretion rate onto neutron stars has been discussed in \citep{kouvaris2008wimp} and the total amount of dark matter accumulated onto a $2\,\mathrm{M_\odot}$ neutron star in a given time $t$ (i.e. a neutron star with age $t$) can be estimated \citep{del2020change, ivanytskyi2020neutron} as
\begin{equation}
    \label{massa di materia oscura in stelle di neutroni}
    \begin{split}
        M_{DM}^{acc} \approx 10^{-14}&\Bigl(\frac{\rho_\chi}{0.3\ \mathrm{GeV/cm^{3}}}\Bigr) \times \\
        &\times \Bigl(\frac{\sigma_{\chi n}}{10^{-45}\ \mathrm{cm^2}}\Bigr)\Bigl(\frac{t}{1\ \mathrm{Gyr}}\Bigr)\ \mathrm{M_\odot}
    \end{split}
\end{equation}
where $\rho_\chi$ is the local dark matter density and $\sigma_{\chi n}$ is the dark matter-nucleon interaction cross section. 

To model the dark matter density profile $\rho_\chi$, we can use the Einasto profile \citep{einasto1965,einasto1969}, since it is in good agreement with numerical simulations and data from the rotation curve of the Milky Way \citep{sofue2013rotation, pato2015dynamical, einasto1965, einasto1969, merritt2006, baes2022}. 
As a function of the distance $r$ from the galactic center we can write
\begin{equation*}
    \frac{\rho_{\chi}(r;\alpha, r_{-2}, \rho_{-2})}{\rho_{-2}} = \exp\Biggl\{-\frac{2}{\alpha}\Biggl[\Bigl(\frac{r}{r_{-2}}\Bigr)^{\alpha}-1\Biggr]\Biggr\}
\end{equation*}
where $\alpha$ is a parameter that determine the curvature of the profile, $\rho_{-2}$ is a normalization constant and $r_{-2}$ is defined by the following condition:
\begin{equation*}
    \frac{d\ln \rho_{\chi}(r)}{d\ln r}\Biggr|_{r = r_{-2}} = -2
\end{equation*}
As discussed in \citep{del2020change}, we can set $r_{-2} = 15\ \mathrm{kpc}$ and fix $\rho_{-2}$ by using the fact that in the solar neighborhood ($r=R_\odot=8\ \mathrm{kpc}$) we should have $\rho_{\chi}(R_\odot)=0.420\ \mathrm{GeV/cm^3}$ \citep{pato2015dynamical}. Depending on the value of the $\alpha$ parameter, we get different density profiles. The total mass of accreted DM onto the neutron star is not very sensitive to the value of the $\alpha$ parameter in the galaxy disc. For a typical neutron star in the solar neighborhood one can get 
$M_{DM}^{acc}\approx 10^{-11}\ \mathrm{M_\odot}$ \citep{del2020change}, while the same estimate for PSR J0348+0432 and PSR J0740+6620 leads to $M_{DM}^{acc} \approx 10^{-13}-10^{-14}\ \mathrm{M_\odot}$ \citep{ivanytskyi2020neutron} depending on the value of the parameters. These values for $M_{DM}^{acc}$ are too low to produce any significant effect on the neutron star structure and to put a stringent upper limit on $m_\chi$. \\
To get an upper limit for the mass of the DM particles, we should consider a very old neutron star ($t\sim 10\ \mathrm{Gyr}$) in a DM rich region like the galactic center. At a distance $r\sim 10^{-5}\ \mathrm{pc}$, we can have up to $M_{DM}^{acc}\approx 10^{-4}-10^{-3}\ \mathrm{M_\odot}$ depending on the value of the parameter $\alpha$ in the Einasto profile \citep{del2020change}. 
In \citep{ivanytskyi2020neutron}, using this approach and taking also into account the DM accretion during the main sequence phase of the neutron star progenitor, a value of $f_\chi = M_{DM}^{acc}/M_{tot} \approx 0.01\%$ is obtained for a $2\ \mathrm{M_\odot}$ neutron star in the galactic center.\\
The latter estimate for $f_\chi$ is represented by the blue dashed line in Fig.\,(\ref{fig: m_fD_2Msun}).  
Combining this latter estimate with our calculations reported in Fig.\,(\ref{fig: m_fD_2Msun}), our equation \eqref{fit} gives us an upper limit for the dark matter particles
\begin{equation}
    \label{DM particle mass upper limit}
    m_\chi \le 56\ \mathrm{GeV}
\end{equation}
which is in good agreement to the result obtained in \citep{ivanytskyi2020neutron} ($m_\chi \le 60\ \mathrm{GeV}$) in spite the fact that we are using a different EOS model for ordinary matter.  

Notice that in the DM accretion rate estimate made in \citep{kouvaris2008wimp}, only the dark matter-nucleon interaction cross-section $\sigma_{\chi n}$ is taken into account. 
If a non zero self-interaction between dark matter particles 
is assumed, the accretion rate 
may change drastically \citep{guver2014capture, nelson2019dark}. This is because self-interactions between DM particles can dominate the capture process when the dark matter-nucleon cross-section $\sigma_{\chi n}$ is very small. \\
Moreover, other sources of DM accretion onto neutron stars should be considered. 
For example, in the central region of the Galaxy, clumps of DM
could interact with the neutron star or its progenitor star leading to a large amount of accreted DM on the star. In \citep{deliyergiyev2024neutron}, without taking into account DM self-interaction, an accretion of $M_{DM}^{acc}\approx 7.5\times 10^{-4}\ \mathrm{M_\odot}$ is estimated for a typical neutron star inside a DM clump.\\
Finally, a significant amount of dark matter could be generated during the supernova explosion, with the majority remaining trapped within the neutron star \citep{nelson2019dark}. 
Taking into account all these contributions and the uncertainties associated with each, one might arrive at a significantly higher estimate ($f_\chi \lesssim 10^{-2}$) of the DM fraction in a neutron star, with respect to the one ($f_\chi \sim 10^{-4}$) mentioned before and reported in Fig.\,(\ref{fig: m_fD_2Msun}). \\ 
The DM fraction values we consider in the present work 
(Fig.\,(\ref{fig: Plot MR fix fchi 1GeV}) and Fig.\,(\ref{fig: Plot MR fix fchi 0.1GeV})) and commonly considered in the literature (e.g.\citep{ellis2018dark,ivanytskyi2020neutron,grippa2024constraintsscalarvectordark}) 
are thus compatible with the upper estimate ($f_\chi \lesssim 10^{-2}$) discussed above.

\section{Stability of DANS} 
\label{sec:stab}

A solution the two-fluid TOV equations \eqref{2fTOV_a} and \eqref{2fTOV_b} represents an equilibrium configuration. In this section we present results regarding the stability with respect to small radial perturbation of DANS. We assume that these perturbations do not modify the $\beta$-equilibrium condition of ordinary matter.

In the one-fluid limit (stars made only of ordinary matter or only of dark matter) it is well known that the critical point is the maximum of the $M(\rho_{c,j})$ curve \citep{Shapiro-Teukolsky_1983}.  
Using the EOSs considered in this work we get the following value for the central mass densities of the critical configuration
\begin{equation*}
    \begin{split}
        \textbf{OM stars:} \quad &\rho_{cOM} = 2.727\times 10^{15}\ \mathrm{g/cm^3}  \\
        \textbf{DM stars:} \quad &\rho_{cDM} = 5.362\times 10^{15}\Bigl(\frac{m_\chi}{1\ \mathrm{GeV}}\Bigr)^4\ \mathrm{g/cm^3} \\
    \end{split}
\end{equation*}

\begin{figure}[!ht]
    \centering
    \includegraphics[width=0.8\linewidth]{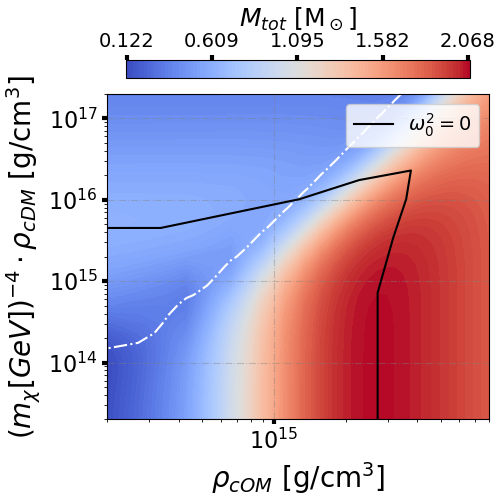}
    \caption{Critical stability curve $\omega_0^2(\rho_{cOM}, \rho_{cDM})=0$ for $m_\chi = 1\ \mathrm{GeV}$ plotted as a contour line in the two central densities plane. Stable DANS configurations ($\omega_0^2>0$) are located in the lower left region of the two density plane, bounded by the black line. The different shades of color denote the function $M_{tot}(\rho_{cOM}, \rho_{cDM})$.\\
    We plot the $\Delta R = R_{OM}-R_{DM}=0$ contour line (white dashed-dotted line) in order to show that both stable DANS with a DM core and DANS with a DM halo are possible solutions to the two-fluid TOV equations (cfr. Fig.\ref{fig: TOV2f plane-integration 1GeV}).}
    \label{fig: stability 1GeV}
\end{figure}

In the two-fluids system case the stability analysis can be done in two different ways. 
The first method, discussed in \citep{kain2020radial}, requires us to solve a set of pulsation equations (eigenvalues problem) whose solutions give the squared radial oscillation frequencies $\omega^2_n(\rho_{cOM}, \rho_{cDM})$ of the different modes. 
For any choice of the central densities of the two fluids $(\rho_{cOM}, \rho_{cDM})$, if the squared radial oscillation frequency for the fundamental mode $\omega_0^2(\rho_{cOM}, \rho_{cDM})$ is positive, then the corresponding solution (DANS) of the two fluids TOV equations is stable with respect to small radial perturbations while, if $\omega_0^2(\rho_{cOM}, \rho_{cDM})<0$, the solution is unstable. The points in the two central density plane in which we have a transition from stable to unstable solutions are defined by the condition $\omega_0^2(\rho_{cOM}, \rho_{cDM}) = 0$: this indicates that in the case of two-fluid systems we have a critical stability curve that corresponds to the geometric locus of points on the plane $(\rho_{cOM}, \rho_{cDM})$ for which $\omega_0^2=0$. 
Determine the critical stability curve with this approach, requires us to take the derivative of the EOSs of the fluids which is a problem if the EOSs aren't smooth enough (as in the case of the BL EOS, which is given in tabular form). \\
Since we are just interested in determining the stability region on the plane of the two central densities, a second method can be used to study the stability of two-fluids systems. This second approach, discussed by \citep{henriques1990stability} for Fermion-Boson stars, allows us to determine the critical stability curve without explicitly solving the pulsation equations. It can be shown that the critical curve is defined by the following set of equations
\begin{equation}
    \label{curva critica}
    \frac{dM_{tot}}{d\vec{\sigma}}=\frac{dN_{OM}}{d\vec{\sigma}}=\frac{dN_{DM}}{d\vec{\sigma}}=0
\end{equation}
where $M_{tot}$, $N_{i}$ ($i=OM, DM$) are the total gravitational mass and total fluid number of a static solution of the two-fluids TOV equations and $\vec{\sigma}$ is a vector in the parameter space (in this case the two central densities plane) that is simultaneously tangent to the level curves of $M_{tot}$ and $N_i$.\\
It can be shown that to determine the critical stability curve, we can study the stationary point of $M_{tot}(\rho_{cOM}, \rho_{cDM})$ along level curve of either $N_{OM}(\rho_{cOM}, \rho_{cDM})$ or $N_{DM}(\rho_{cOM}, \rho_{cDM})$ (see also \citep{valdez2013dynamical, gleason2022dynamical}). As observed in \citep{henriques1990stability}, the stability region must contain the stable structures in the one-fluid limit. \\
We apply the latter method in the $m_\chi=1\ \mathrm{GeV}$ case (see results in Fig.\eqref{fig: stability 1GeV}) by studying stationary points of $M_{tot}(\rho_{cOM}, \rho_{cDM})$ along $N_{DM}(\rho_{cOM}, \rho_{cDM})$ level curve, in order to check that in our model we get a reasonably large stability region in the central densities plane. \\
Our results are in agreement with those obtained in \citep{kain2021dark}, and show us that we can have stable DANS with both a DM core or a DM halo (cfr central panel of Fig \ref{fig: TOV2f plane-integration 1GeV}) and that exists a region inside the stability curve where DANS have an ordinary matter central density \textit{and/or} a dark matter central density that exceed the critical value in the one-fluid limit.

\section{Summary} 
\label{sec:conclusions}
In this paper we have studied the consequences of the presence of DM on the structural properties and the stability of neutron stars. 
Specifically, we explored how the properties of these DANS are modified with respect to those of ordinary neutron stars in terms of the DM particle 
mass $m_\chi$ and the fraction of DM present in the star.  
To model ordinary matter we used a state-of-the-art equation of state of $\beta$-stable nuclear matter obtained using the Brueckner-Hartree-Fock quantum many-body approach starting from two-body and three-body nuclear interactions derived from chiral effective field theory.  
This EOS, as discussed in Sect.\,\ref{sec:BLeos}, reproduces various empirical nuclear matter properties at the saturation density $n_0$ and at supranuclear densities ($n >\, n_0$). Additionally, the calculated $E_{sym}(n_0)$ and $L$ fulfill the bounds imposed by the unitary Fermi gas limit \citep{Tews2017}   
and are compatible with the values extracted in Ref.~\citep{PhysRevLett.126.172503} from the measured neutron skin thickness of the $^{208}$Pb nucleus in the PREX-2 experiment \citep{PREX2}.  
The dark matter component of the star has been modeled as a non-self-annihilating system of spin $1/2$ fermions and its EOS as an ideal relativistic Fermi gas.  We found that, depending on the dark matter particle mass $m_\chi$, one can have somehow opposite effects on the stellar properties 
when a DM fraction $f_\chi$ is added to the stellar material.  
In fact, in the case $m_\chi = 1\, \mathrm{GeV}$, the stellar gravitational maximum mass $M_{max}$ decreases, whereas in the case 
$m_\chi = 0.1\, \mathrm{GeV}$ $M_{max}$ increases with respect to the maximum mass of ordinary neutron stars.   
We studied $M_{max}$ as a function of $f_\chi$ for some fixed values of the DM particle mass $m_\chi$.  
We defined the critical values $f_\chi^{crit}(m_\chi)$ as the roots of the equation $M(f_\chi, m_\chi) = 2\ \mathrm{M_\odot}$ for a given value of $m_\chi$. Comparing an upper limit for the total dark matter fraction accumulated by a $2\ \mathrm{M_\odot}$ neutron star in the galactic center with the function $f_\chi^{crit}(m_\chi)$, we found an upper limit for the dark matter particles: $m_\chi \le 56\ \mathrm{GeV}$.  
We also shown that the presence of dark matter has indirect sizable effect on the proton fraction in the ordinary matter fluid and, in the case $m_\chi = 1\, \mathrm{GeV}$, results in a decrease of the threshold gravitational mass $M_{tot}^{durca}$ for having direct URCA processes and fast stellar cooling. 

\bibliographystyle{elsarticle-harv} 
\bibliography{biblio}

\begin{thebibliography}{85}
\expandafter\ifx\csname natexlab\endcsname\relax\def\natexlab#1{#1}\fi
\providecommand{\url}[1]{\texttt{#1}}
\providecommand{\href}[2]{#2}
\providecommand{\path}[1]{#1}
\providecommand{\DOIprefix}{doi:}
\providecommand{\ArXivprefix}{arXiv:}
\providecommand{\URLprefix}{URL: }
\providecommand{\Pubmedprefix}{pmid:}
\providecommand{\doi}[1]{\href{http://dx.doi.org/#1}{\path{#1}}}
\providecommand{\Pubmed}[1]{\href{pmid:#1}{\path{#1}}}
\providecommand{\bibinfo}[2]{#2}
\ifx\xfnm\relax \def\xfnm[#1]{\unskip,\space#1}\fi
\bibitem[{Abbott et~al.(2017)}]{TheLIGOScientific:2017qsa}
\bibinfo{author}{Abbott, B.P.}, et~al. (\bibinfo{collaboration}{LIGO
  Scientific, Virgo}), \bibinfo{year}{2017}.
\newblock \bibinfo{title}{{GW170817: Observation of Gravitational Waves from a
  Binary Neutron Star Inspiral}}.
\newblock \bibinfo{journal}{Phys. Rev. Lett.} \bibinfo{volume}{119},
  \bibinfo{pages}{161101}.
\newblock \DOIprefix\doi{10.1103/PhysRevLett.119.161101},
  \href{http://arxiv.org/abs/1710.05832}{{\tt arXiv:1710.05832}}.
\bibitem[{Abbott et~al.(2018)}]{GW170817_MR}
\bibinfo{author}{Abbott, B.P.}, et~al. (\bibinfo{collaboration}{The LIGO
  Scientific Collaboration and the Virgo Collaboration}), \bibinfo{year}{2018}.
\newblock \bibinfo{title}{Gw170817: Measurements of neutron star radii and
  equation of state}.
\newblock \bibinfo{journal}{Phys. Rev. Lett.} \bibinfo{volume}{121},
  \bibinfo{pages}{161101}.
\newblock \URLprefix
  \url{https://link.aps.org/doi/10.1103/PhysRevLett.121.161101},
  \DOIprefix\doi{10.1103/PhysRevLett.121.161101}.
\bibitem[{Adhikari et~al.(2021)Adhikari, Albataineh, Androic, Aniol, Armstrong,
  Averett, Ayerbe~Gayoso, Barcus, Bellini, Beminiwattha, Benesch, Bhatt,
  Bhatta~Pathak, Bhetuwal, Blaikie, Campagna, Camsonne, Cates, Chen, Clarke,
  Cornejo, Covrig~Dusa, Datta, Deshpande, Dutta, Feldman, Fuchey, Gal, Gaskell,
  Gautam, Gericke, Ghosh, Halilovic, Hansen, Hauenstein, Henry, Horowitz,
  Jantzi, Jian, Johnston, Jones, Karki, Katugampola, Keppel, King, King,
  Knauss, Kumar, Kutz, Lashley-Colthirst, Leverick, Liu, Liyange, Malace,
  Mammei, Mammei, McCaughan, McNulty, Meekins, Metts, Michaels, Mondal,
  Napolitano, Narayan, Nikolaev, Rashad, Owen, Palatchi, Pan, Pandey, Park,
  Paschke, Petrusky, Pitt, Premathilake, Puckett, Quinn, Radloff, Rahman,
  Rathnayake, Reed, Reimer, Richards, Riordan, Roblin, Seeds, Shahinyan,
  Souder, Tang, Thiel, Tian, Urciuoli, Wertz, Wojtsekhowski, Yale, Ye, Yoon,
  Zec, Zhang, Zhang and Zheng}]{PREX2}
\bibinfo{author}{Adhikari, D.}, \bibinfo{author}{Albataineh, H.},
  \bibinfo{author}{Androic, D.}, \bibinfo{author}{Aniol, K.},
  \bibinfo{author}{Armstrong, D.S.}, \bibinfo{author}{Averett, T.},
  \bibinfo{author}{Ayerbe~Gayoso, C.}, \bibinfo{author}{Barcus, S.},
  \bibinfo{author}{Bellini, V.}, \bibinfo{author}{Beminiwattha, R.S.},
  \bibinfo{author}{Benesch, J.F.}, \bibinfo{author}{Bhatt, H.},
  \bibinfo{author}{Bhatta~Pathak, D.}, \bibinfo{author}{Bhetuwal, D.},
  \bibinfo{author}{Blaikie, B.}, \bibinfo{author}{Campagna, Q.},
  \bibinfo{author}{Camsonne, A.}, \bibinfo{author}{Cates, G.D.},
  \bibinfo{author}{Chen, Y.}, \bibinfo{author}{Clarke, C.},
  \bibinfo{author}{Cornejo, J.C.}, \bibinfo{author}{Covrig~Dusa, S.},
  \bibinfo{author}{Datta, P.}, \bibinfo{author}{Deshpande, A.},
  \bibinfo{author}{Dutta, D.}, \bibinfo{author}{Feldman, C.},
  \bibinfo{author}{Fuchey, E.}, \bibinfo{author}{Gal, C.},
  \bibinfo{author}{Gaskell, D.}, \bibinfo{author}{Gautam, T.},
  \bibinfo{author}{Gericke, M.}, \bibinfo{author}{Ghosh, C.},
  \bibinfo{author}{Halilovic, I.}, \bibinfo{author}{Hansen, J.O.},
  \bibinfo{author}{Hauenstein, F.}, \bibinfo{author}{Henry, W.},
  \bibinfo{author}{Horowitz, C.J.}, \bibinfo{author}{Jantzi, C.},
  \bibinfo{author}{Jian, S.}, \bibinfo{author}{Johnston, S.},
  \bibinfo{author}{Jones, D.C.}, \bibinfo{author}{Karki, B.},
  \bibinfo{author}{Katugampola, S.}, \bibinfo{author}{Keppel, C.},
  \bibinfo{author}{King, P.M.}, \bibinfo{author}{King, D.E.},
  \bibinfo{author}{Knauss, M.}, \bibinfo{author}{Kumar, K.S.},
  \bibinfo{author}{Kutz, T.}, \bibinfo{author}{Lashley-Colthirst, N.},
  \bibinfo{author}{Leverick, G.}, \bibinfo{author}{Liu, H.},
  \bibinfo{author}{Liyange, N.}, \bibinfo{author}{Malace, S.},
  \bibinfo{author}{Mammei, R.}, \bibinfo{author}{Mammei, J.},
  \bibinfo{author}{McCaughan, M.}, \bibinfo{author}{McNulty, D.},
  \bibinfo{author}{Meekins, D.}, \bibinfo{author}{Metts, C.},
  \bibinfo{author}{Michaels, R.}, \bibinfo{author}{Mondal, M.M.},
  \bibinfo{author}{Napolitano, J.}, \bibinfo{author}{Narayan, A.},
  \bibinfo{author}{Nikolaev, D.}, \bibinfo{author}{Rashad, M.N.H.},
  \bibinfo{author}{Owen, V.}, \bibinfo{author}{Palatchi, C.},
  \bibinfo{author}{Pan, J.}, \bibinfo{author}{Pandey, B.},
  \bibinfo{author}{Park, S.}, \bibinfo{author}{Paschke, K.D.},
  \bibinfo{author}{Petrusky, M.}, \bibinfo{author}{Pitt, M.L.},
  \bibinfo{author}{Premathilake, S.}, \bibinfo{author}{Puckett, A.J.R.},
  \bibinfo{author}{Quinn, B.}, \bibinfo{author}{Radloff, R.},
  \bibinfo{author}{Rahman, S.}, \bibinfo{author}{Rathnayake, A.},
  \bibinfo{author}{Reed, B.T.}, \bibinfo{author}{Reimer, P.E.},
  \bibinfo{author}{Richards, R.}, \bibinfo{author}{Riordan, S.},
  \bibinfo{author}{Roblin, Y.}, \bibinfo{author}{Seeds, S.},
  \bibinfo{author}{Shahinyan, A.}, \bibinfo{author}{Souder, P.},
  \bibinfo{author}{Tang, L.}, \bibinfo{author}{Thiel, M.},
  \bibinfo{author}{Tian, Y.}, \bibinfo{author}{Urciuoli, G.M.},
  \bibinfo{author}{Wertz, E.W.}, \bibinfo{author}{Wojtsekhowski, B.},
  \bibinfo{author}{Yale, B.}, \bibinfo{author}{Ye, T.}, \bibinfo{author}{Yoon,
  A.}, \bibinfo{author}{Zec, A.}, \bibinfo{author}{Zhang, W.},
  \bibinfo{author}{Zhang, J.}, \bibinfo{author}{Zheng, X.}
  (\bibinfo{collaboration}{PREX Collaboration}), \bibinfo{year}{2021}.
\newblock \bibinfo{title}{Accurate determination of the neutron skin thickness
  of $^{208}\mathrm{Pb}$ through parity-violation in electron scattering}.
\newblock \bibinfo{journal}{Phys. Rev. Lett.} \bibinfo{volume}{126},
  \bibinfo{pages}{172502}.
\newblock \URLprefix
  \url{https://link.aps.org/doi/10.1103/PhysRevLett.126.172502},
  \DOIprefix\doi{10.1103/PhysRevLett.126.172502}.
\bibitem[{Akmal et~al.(1998)Akmal, Pandharipande and Ravenhall}]{Akmal:1998cf}
\bibinfo{author}{Akmal, A.}, \bibinfo{author}{Pandharipande, V.R.},
  \bibinfo{author}{Ravenhall, D.G.}, \bibinfo{year}{1998}.
\newblock \bibinfo{title}{{The Equation of state of nucleon matter and neutron
  star structure}}.
\newblock \bibinfo{journal}{Phys. Rev. C} \bibinfo{volume}{58},
  \bibinfo{pages}{1804--1828}.
\newblock \DOIprefix\doi{10.1103/PhysRevC.58.1804},
  \href{http://arxiv.org/abs/nucl-th/9804027}{{\tt arXiv:nucl-th/9804027}}.
\bibitem[{{Antoniadis} et~al.(2013){Antoniadis}, {Freire}, {Wex}, {Tauris},
  {Lynch}, {van Kerkwijk}, {Kramer}, {Bassa}, {Dhillon}, {Driebe}, {Hessels},
  {Kaspi}, {Kondratiev}, {Langer}, {Marsh}, {McLaughlin}, {Pennucci}, {Ransom},
  {Stairs}, {van Leeuwen}, {Verbiest} and {Whelan}}]{antoniadis2013massive}
\bibinfo{author}{{Antoniadis}, J.}, \bibinfo{author}{{Freire}, P.C.C.},
  \bibinfo{author}{{Wex}, N.}, \bibinfo{author}{{Tauris}, T.M.},
  \bibinfo{author}{{Lynch}, R.S.}, \bibinfo{author}{{van Kerkwijk}, M.H.},
  \bibinfo{author}{{Kramer}, M.}, \bibinfo{author}{{Bassa}, C.},
  \bibinfo{author}{{Dhillon}, V.S.}, \bibinfo{author}{{Driebe}, T.},
  \bibinfo{author}{{Hessels}, J.W.T.}, \bibinfo{author}{{Kaspi}, V.M.},
  \bibinfo{author}{{Kondratiev}, V.I.}, \bibinfo{author}{{Langer}, N.},
  \bibinfo{author}{{Marsh}, T.R.}, \bibinfo{author}{{McLaughlin}, M.A.},
  \bibinfo{author}{{Pennucci}, T.T.}, \bibinfo{author}{{Ransom}, S.M.},
  \bibinfo{author}{{Stairs}, I.H.}, \bibinfo{author}{{van Leeuwen}, J.},
  \bibinfo{author}{{Verbiest}, J.P.W.}, \bibinfo{author}{{Whelan}, D.G.},
  \bibinfo{year}{2013}.
\newblock \bibinfo{title}{{A Massive Pulsar in a Compact Relativistic Binary}}.
\newblock \bibinfo{journal}{Science} \bibinfo{volume}{340},
  \bibinfo{pages}{448}.
\newblock \DOIprefix\doi{10.1126/science.1233232},
  \href{http://arxiv.org/abs/1304.6875}{{\tt arXiv:1304.6875}}.
\bibitem[{Aprile et~al.(2017)Aprile, Aalbers, Agostini, Alfonsi, Amaro,
  Anthony, Arneodo, Barrow, Baudis, Bauermeister et~al.}]{aprile2017first}
\bibinfo{author}{Aprile, E.}, \bibinfo{author}{Aalbers, J.},
  \bibinfo{author}{Agostini, F.}, \bibinfo{author}{Alfonsi, M.},
  \bibinfo{author}{Amaro, F.}, \bibinfo{author}{Anthony, M.},
  \bibinfo{author}{Arneodo, F.}, \bibinfo{author}{Barrow, P.},
  \bibinfo{author}{Baudis, L.}, \bibinfo{author}{Bauermeister, B.}, et~al.,
  \bibinfo{year}{2017}.
\newblock \bibinfo{title}{First dark matter search results from the xenon1t
  experiment}.
\newblock \bibinfo{journal}{Physical Review Letters} \bibinfo{volume}{119},
  \bibinfo{pages}{181301}.
\bibitem[{{\'A}vila et~al.(2024){\'A}vila, Giangrandi, Sagun, Ivanytskyi and
  Provid{\^e}ncia}]{avila2024rapid}
\bibinfo{author}{{\'A}vila, A.}, \bibinfo{author}{Giangrandi, E.},
  \bibinfo{author}{Sagun, V.}, \bibinfo{author}{Ivanytskyi, O.},
  \bibinfo{author}{Provid{\^e}ncia, C.}, \bibinfo{year}{2024}.
\newblock \bibinfo{title}{Rapid neutron star cooling triggered by dark matter}.
\newblock \bibinfo{journal}{Monthly Notices of the Royal Astronomical Society}
  \bibinfo{volume}{528}, \bibinfo{pages}{6319--6328}.
\bibitem[{{Baes}(2022)}]{baes2022}
\bibinfo{author}{{Baes}, M.}, \bibinfo{year}{2022}.
\newblock \bibinfo{title}{{The Einasto model for dark matter haloes}}.
\newblock \bibinfo{journal}{Astronomy \& Astrophysics} \bibinfo{volume}{667},
  \bibinfo{pages}{A47}.
\newblock \DOIprefix\doi{10.1051/0004-6361/202244567},
  \href{http://arxiv.org/abs/2209.03639}{{\tt arXiv:2209.03639}}.
\bibitem[{Barbat et~al.(2024)Barbat, Schaffner-Bielich and
  Tolos}]{Barbat:2024yvi}
\bibinfo{author}{Barbat, M.F.}, \bibinfo{author}{Schaffner-Bielich, J.},
  \bibinfo{author}{Tolos, L.}, \bibinfo{year}{2024}.
\newblock \bibinfo{title}{{Comprehensive study of compact stars with dark
  matter}}.
\newblock \bibinfo{journal}{Phys. Rev. D} \bibinfo{volume}{110},
  \bibinfo{pages}{023013}.
\newblock \DOIprefix\doi{10.1103/PhysRevD.110.023013},
  \href{http://arxiv.org/abs/2404.12875}{{\tt arXiv:2404.12875}}.
\bibitem[{{Baym} et~al.(1971){Baym}, {Pethick} and {Sutherland}}]{bps71}
\bibinfo{author}{{Baym}, G.}, \bibinfo{author}{{Pethick}, C.},
  \bibinfo{author}{{Sutherland}, P.}, \bibinfo{year}{1971}.
\newblock \bibinfo{title}{{The Ground State of Matter at High Densities:
  Equation of State and Stellar Models}}.
\newblock \bibinfo{journal}{Astrophys. J.} \bibinfo{volume}{170},
  \bibinfo{pages}{299}.
\newblock \DOIprefix\doi{10.1086/151216}.
\bibitem[{Bernuzzi et~al.(2020)}]{Bernuzzi:2020txg}
\bibinfo{author}{Bernuzzi, S.}, et~al., \bibinfo{year}{2020}.
\newblock \bibinfo{title}{{Accretion-induced prompt black hole formation in
  asymmetric neutron star mergers, dynamical ejecta and kilonova signals}}.
\newblock \bibinfo{journal}{Mon. Not. Roy. Astron. Soc.} \bibinfo{volume}{497},
  \bibinfo{pages}{1488--1507}.
\newblock \DOIprefix\doi{10.1093/mnras/staa1860},
  \href{http://arxiv.org/abs/2003.06015}{{\tt arXiv:2003.06015}}.
\bibitem[{Bertone and Fairbairn(2008)}]{bertone2008compact}
\bibinfo{author}{Bertone, G.}, \bibinfo{author}{Fairbairn, M.},
  \bibinfo{year}{2008}.
\newblock \bibinfo{title}{Compact stars as dark matter probes}.
\newblock \bibinfo{journal}{Physical Review D} \bibinfo{volume}{77},
  \bibinfo{pages}{043515}.
\bibitem[{Bertone et~al.(2005)Bertone, Hooper and Silk}]{bertone2005particle}
\bibinfo{author}{Bertone, G.}, \bibinfo{author}{Hooper, D.},
  \bibinfo{author}{Silk, J.}, \bibinfo{year}{2005}.
\newblock \bibinfo{title}{Particle dark matter: Evidence, candidates and
  constraints}.
\newblock \bibinfo{journal}{Physics Reports} \bibinfo{volume}{405},
  \bibinfo{pages}{279--390}.
\bibitem[{{Bombaci}(2022)}]{Bombaci_book_2022}
\bibinfo{author}{{Bombaci}, I.}, \bibinfo{year}{2022}.
\newblock \bibinfo{title}{{The Equation of State of Neutron Star Matter}}, in:
  \bibinfo{editor}{{Bhattacharyya}, S.}, \bibinfo{editor}{{Papitto}, A.},
  \bibinfo{editor}{{Bhattacharya}, D.} (Eds.), \bibinfo{booktitle}{Astrophysics
  and Space Science Library}, pp. \bibinfo{pages}{281--317}.
\newblock \DOIprefix\doi{10.1007/978-3-030-85198-9_9}.
\bibitem[{Bombaci and Logoteta(2018)}]{BL2018}
\bibinfo{author}{Bombaci, I.}, \bibinfo{author}{Logoteta, D.},
  \bibinfo{year}{2018}.
\newblock \bibinfo{title}{{Equation of state of dense nuclear matter and
  neutron star structure from nuclear chiral interactions}}.
\newblock \bibinfo{journal}{Astron. Astrophys.} \bibinfo{volume}{609},
  \bibinfo{pages}{A128}.
\newblock \DOIprefix\doi{10.1051/0004-6361/201731604},
  \href{http://arxiv.org/abs/1805.11846}{{\tt arXiv:1805.11846}}.
\bibitem[{{Bombaci} and {Lombardo}(1991)}]{BL_1991}
\bibinfo{author}{{Bombaci}, I.}, \bibinfo{author}{{Lombardo}, U.},
  \bibinfo{year}{1991}.
\newblock \bibinfo{title}{{Asymmetric nuclear matter equation of state}}.
\newblock \bibinfo{journal}{Phys. Rev. C} \bibinfo{volume}{44},
  \bibinfo{pages}{1892--1900}.
\newblock \DOIprefix\doi{10.1103/PhysRevC.44.1892}.
\bibitem[{{Ciarcelluti} and {Sandin}(2011)}]{Ciarcelluti2011}
\bibinfo{author}{{Ciarcelluti}, P.}, \bibinfo{author}{{Sandin}, F.},
  \bibinfo{year}{2011}.
\newblock \bibinfo{title}{{Have neutron stars a dark matter core?}}
\newblock \bibinfo{journal}{Physics Letters B} \bibinfo{volume}{695},
  \bibinfo{pages}{19--21}.
\newblock \DOIprefix\doi{10.1016/j.physletb.2010.11.021},
  \href{http://arxiv.org/abs/1005.0857}{{\tt arXiv:1005.0857}}.
\bibitem[{Corbelli and Salucci(2000)}]{corbelli2000extended}
\bibinfo{author}{Corbelli, E.}, \bibinfo{author}{Salucci, P.},
  \bibinfo{year}{2000}.
\newblock \bibinfo{title}{The extended rotation curve and the dark matter halo
  of m33}.
\newblock \bibinfo{journal}{Monthly Notices of the Royal Astronomical Society}
  \bibinfo{volume}{311}, \bibinfo{pages}{441--447}.
\bibitem[{Danielewicz et~al.(2002)Danielewicz, Lacey and
  Lynch}]{Danielewicz:2002pu}
\bibinfo{author}{Danielewicz, P.}, \bibinfo{author}{Lacey, R.},
  \bibinfo{author}{Lynch, W.G.}, \bibinfo{year}{2002}.
\newblock \bibinfo{title}{{Determination of the equation of state of dense
  matter}}.
\newblock \bibinfo{journal}{Science} \bibinfo{volume}{298},
  \bibinfo{pages}{1592--1596}.
\newblock \DOIprefix\doi{10.1126/science.1078070},
  \href{http://arxiv.org/abs/nucl-th/0208016}{{\tt arXiv:nucl-th/0208016}}.
\bibitem[{Danielewicz and Lee(2014)}]{danielewicz2014symmetry}
\bibinfo{author}{Danielewicz, P.}, \bibinfo{author}{Lee, J.},
  \bibinfo{year}{2014}.
\newblock \bibinfo{title}{Symmetry energy ii: Isobaric analog states}.
\newblock \bibinfo{journal}{Nuclear Physics A} \bibinfo{volume}{922},
  \bibinfo{pages}{1--70}.
\bibitem[{Del~Popolo et~al.(2020)Del~Popolo, Deliyergiyev, Le~Delliou, Tolos
  and Burgio}]{del2020change}
\bibinfo{author}{Del~Popolo, A.}, \bibinfo{author}{Deliyergiyev, M.},
  \bibinfo{author}{Le~Delliou, M.}, \bibinfo{author}{Tolos, L.},
  \bibinfo{author}{Burgio, F.}, \bibinfo{year}{2020}.
\newblock \bibinfo{title}{On the change of old neutron star masses with
  galactocentric distance}.
\newblock \bibinfo{journal}{Physics of the Dark Universe} \bibinfo{volume}{28},
  \bibinfo{pages}{100484}.
\bibitem[{Deliyergiyev et~al.(2024)Deliyergiyev, Del~Popolo and
  Le~Delliou}]{deliyergiyev2024neutron}
\bibinfo{author}{Deliyergiyev, M.}, \bibinfo{author}{Del~Popolo, A.},
  \bibinfo{author}{Le~Delliou, M.}, \bibinfo{year}{2024}.
\newblock \bibinfo{title}{Neutron star mass in dark matter clumps}.
\newblock \bibinfo{journal}{Monthly Notices of the Royal Astronomical Society}
  \bibinfo{volume}{527}, \bibinfo{pages}{4483--4504}.
\bibitem[{Deliyergiyev et~al.(2019)Deliyergiyev, Del~Popolo, Tolos, Le~Delliou,
  Lee and Burgio}]{deliyergiyev2019dark}
\bibinfo{author}{Deliyergiyev, M.}, \bibinfo{author}{Del~Popolo, A.},
  \bibinfo{author}{Tolos, L.}, \bibinfo{author}{Le~Delliou, M.},
  \bibinfo{author}{Lee, X.}, \bibinfo{author}{Burgio, F.},
  \bibinfo{year}{2019}.
\newblock \bibinfo{title}{Dark compact objects: an extensive overview}.
\newblock \bibinfo{journal}{Physical Review D} \bibinfo{volume}{99},
  \bibinfo{pages}{063015}.
\bibitem[{Dengler et~al.(2022)Dengler, Schaffner-Bielich and
  Tolos}]{Dengler:2021qcq}
\bibinfo{author}{Dengler, Y.}, \bibinfo{author}{Schaffner-Bielich, J.},
  \bibinfo{author}{Tolos, L.}, \bibinfo{year}{2022}.
\newblock \bibinfo{title}{{Second Love number of dark compact planets and
  neutron stars with dark matter}}.
\newblock \bibinfo{journal}{Phys. Rev. D} \bibinfo{volume}{105},
  \bibinfo{pages}{043013}.
\newblock \DOIprefix\doi{10.1103/PhysRevD.105.043013},
  \href{http://arxiv.org/abs/2111.06197}{{\tt arXiv:2111.06197}}.
\bibitem[{{Dutra} et~al.(2012){Dutra}, {Louren{\c{c}}o}, {S{\'a} Martins},
  {Delfino}, {Stone} and {Stevenson}}]{Dutra2012}
\bibinfo{author}{{Dutra}, M.}, \bibinfo{author}{{Louren{\c{c}}o}, O.},
  \bibinfo{author}{{S{\'a} Martins}, J.S.}, \bibinfo{author}{{Delfino}, A.},
  \bibinfo{author}{{Stone}, J.R.}, \bibinfo{author}{{Stevenson}, P.D.},
  \bibinfo{year}{2012}.
\newblock \bibinfo{title}{{Skyrme interaction and nuclear matter constraints}}.
\newblock \bibinfo{journal}{prc} \bibinfo{volume}{85}, \bibinfo{pages}{035201}.
\newblock \DOIprefix\doi{10.1103/PhysRevC.85.035201},
  \href{http://arxiv.org/abs/1202.3902}{{\tt arXiv:1202.3902}}.
\bibitem[{{Einasto}(1965)}]{einasto1965}
\bibinfo{author}{{Einasto}, J.}, \bibinfo{year}{1965}.
\newblock \bibinfo{title}{{On the Construction of a Composite Model for the
  Galaxy and on the Determination of the System of Galactic Parameters}}.
\newblock \bibinfo{journal}{Trudy Astrofizicheskogo Instituta Alma-Ata}
  \bibinfo{volume}{5}, \bibinfo{pages}{87--100}.
\bibitem[{{Einasto}(1969)}]{einasto1969}
\bibinfo{author}{{Einasto}, J.}, \bibinfo{year}{1969}.
\newblock \bibinfo{title}{{The Andromeda galaxy M31. I. A preliminary model}}.
\newblock \bibinfo{journal}{Astrofizika} \bibinfo{volume}{5},
  \bibinfo{pages}{137--159}.
\bibitem[{Ellis et~al.(2018)Ellis, H{\"u}tsi, Kannike, Marzola, Raidal and
  Vaskonen}]{ellis2018dark}
\bibinfo{author}{Ellis, J.}, \bibinfo{author}{H{\"u}tsi, G.},
  \bibinfo{author}{Kannike, K.}, \bibinfo{author}{Marzola, L.},
  \bibinfo{author}{Raidal, M.}, \bibinfo{author}{Vaskonen, V.},
  \bibinfo{year}{2018}.
\newblock \bibinfo{title}{Dark matter effects on neutron star properties}.
\newblock \bibinfo{journal}{Physical Review D} \bibinfo{volume}{97},
  \bibinfo{pages}{123007}.
\bibitem[{Endrizzi et~al.(2018)Endrizzi, Logoteta, Giacomazzo, Bombaci, Kastaun
  and Ciolfi}]{Endrizzi:2018uwl}
\bibinfo{author}{Endrizzi, A.}, \bibinfo{author}{Logoteta, D.},
  \bibinfo{author}{Giacomazzo, B.}, \bibinfo{author}{Bombaci, I.},
  \bibinfo{author}{Kastaun, W.}, \bibinfo{author}{Ciolfi, R.},
  \bibinfo{year}{2018}.
\newblock \bibinfo{title}{{Effects of Chiral Effective Field Theory Equation of
  State on Binary Neutron Star Mergers}}.
\newblock \bibinfo{journal}{Phys. Rev. D} \bibinfo{volume}{98},
  \bibinfo{pages}{043015}.
\newblock \DOIprefix\doi{10.1103/PhysRevD.98.043015},
  \href{http://arxiv.org/abs/1806.09832}{{\tt arXiv:1806.09832}}.
\bibitem[{Fonseca et~al.(2021)}]{Fonseca:2021wxt}
\bibinfo{author}{Fonseca, E.}, et~al., \bibinfo{year}{2021}.
\newblock \bibinfo{title}{{Refined Mass and Geometric Measurements of the
  High-mass PSR J0740+6620}}.
\newblock \bibinfo{journal}{Astrophys. J. Lett.} \bibinfo{volume}{915},
  \bibinfo{pages}{L12}.
\newblock \DOIprefix\doi{10.3847/2041-8213/ac03b8},
  \href{http://arxiv.org/abs/2104.00880}{{\tt arXiv:2104.00880}}.
\bibitem[{Garnavich et~al.(1998)Garnavich, Kirshner, Challis, Tonry, Gilliland,
  Smith, Clocchiatti, Diercks, Filippenko, Hamuy, Hogan, Leibundgut, Phillips,
  Reiss, Riess, Schmidt, Schommer, Spyromilio, Stubbs, Suntzeff and
  Wells}]{Garnavich_1998}
\bibinfo{author}{Garnavich, P.M.}, \bibinfo{author}{Kirshner, R.P.},
  \bibinfo{author}{Challis, P.}, \bibinfo{author}{Tonry, J.},
  \bibinfo{author}{Gilliland, R.L.}, \bibinfo{author}{Smith, R.C.},
  \bibinfo{author}{Clocchiatti, A.}, \bibinfo{author}{Diercks, A.},
  \bibinfo{author}{Filippenko, A.V.}, \bibinfo{author}{Hamuy, M.},
  \bibinfo{author}{Hogan, C.J.}, \bibinfo{author}{Leibundgut, B.},
  \bibinfo{author}{Phillips, M.M.}, \bibinfo{author}{Reiss, D.},
  \bibinfo{author}{Riess, A.G.}, \bibinfo{author}{Schmidt, B.P.},
  \bibinfo{author}{Schommer, R.A.}, \bibinfo{author}{Spyromilio, J.},
  \bibinfo{author}{Stubbs, C.}, \bibinfo{author}{Suntzeff, N.B.},
  \bibinfo{author}{Wells, L.}, \bibinfo{year}{1998}.
\newblock \bibinfo{title}{Constraints on cosmological models from hubble space
  telescope observations of high-z supernovae}.
\newblock \bibinfo{journal}{Astrophysical Journal} \bibinfo{volume}{493},
  \bibinfo{pages}{L53}.
\newblock \URLprefix \url{https://dx.doi.org/10.1086/311140},
  \DOIprefix\doi{10.1086/311140}.
\bibitem[{Giangrandi et~al.(2024)Giangrandi, {\'A}vila, Sagun, Ivanytskyi and
  Provid{\^e}ncia}]{giangrandi2024impact}
\bibinfo{author}{Giangrandi, E.}, \bibinfo{author}{{\'A}vila, A.},
  \bibinfo{author}{Sagun, V.}, \bibinfo{author}{Ivanytskyi, O.},
  \bibinfo{author}{Provid{\^e}ncia, C.}, \bibinfo{year}{2024}.
\newblock \bibinfo{title}{The impact of asymmetric dark matter on the thermal
  evolution of nucleonic and hyperonic compact stars}.
\newblock \bibinfo{journal}{Particles} \bibinfo{volume}{7},
  \bibinfo{pages}{179--200}.
\bibitem[{Gleason et~al.(2022)Gleason, Brown and Kain}]{gleason2022dynamical}
\bibinfo{author}{Gleason, T.}, \bibinfo{author}{Brown, B.},
  \bibinfo{author}{Kain, B.}, \bibinfo{year}{2022}.
\newblock \bibinfo{title}{Dynamical evolution of dark matter admixed neutron
  stars}.
\newblock \bibinfo{journal}{Physical Review D} \bibinfo{volume}{105},
  \bibinfo{pages}{023010}.
\bibitem[{{Goldhaber} and {Perlmutter}(1998)}]{GP1998}
\bibinfo{author}{{Goldhaber}, G.}, \bibinfo{author}{{Perlmutter}, S.},
  \bibinfo{year}{1998}.
\newblock \bibinfo{title}{{A study of 42 Type Ia supernovae and a resulting
  measurement of {\ensuremath{\Omega}}$_{M}$ and
  {\ensuremath{\Omega}}$_{{\ensuremath{\Lambda}}}$.}}
\newblock \bibinfo{journal}{Physics Reports} \bibinfo{volume}{307},
  \bibinfo{pages}{325--331}.
\newblock \DOIprefix\doi{10.1016/S0370-1573(98)00091-X}.
\bibitem[{Gresham and Zurek(2019)}]{gresham2019asymmetric}
\bibinfo{author}{Gresham, M.I.}, \bibinfo{author}{Zurek, K.M.},
  \bibinfo{year}{2019}.
\newblock \bibinfo{title}{Asymmetric dark stars and neutron star stability}.
\newblock \bibinfo{journal}{Physical Review D} \bibinfo{volume}{99},
  \bibinfo{pages}{083008}.
\bibitem[{Grippa et~al.(2024)Grippa, Lambiase and
  Poddar}]{grippa2024constraintsscalarvectordark}
\bibinfo{author}{Grippa, F.}, \bibinfo{author}{Lambiase, G.},
  \bibinfo{author}{Poddar, T.K.}, \bibinfo{year}{2024}.
\newblock \bibinfo{title}{Constraints on scalar and vector dark matter admixed
  neutron stars with linear and quadratic couplings}.
\newblock \URLprefix \url{https://arxiv.org/abs/2407.16386},
  \href{http://arxiv.org/abs/2407.16386}{{\tt arXiv:2407.16386}}.
\bibitem[{G{\"u}ver et~al.(2014)G{\"u}ver, Erkoca, Reno and
  Sarcevic}]{guver2014capture}
\bibinfo{author}{G{\"u}ver, T.}, \bibinfo{author}{Erkoca, A.E.},
  \bibinfo{author}{Reno, M.H.}, \bibinfo{author}{Sarcevic, I.},
  \bibinfo{year}{2014}.
\newblock \bibinfo{title}{On the capture of dark matter by neutron stars}.
\newblock \bibinfo{journal}{Journal of Cosmology and Astroparticle Physics}
  \bibinfo{volume}{2014}, \bibinfo{pages}{013}.
\bibitem[{{Hammer} et~al.(2020){Hammer}, {K{\"o}nig} and {van
  Kolck}}]{2020RvMP...92b5004H}
\bibinfo{author}{{Hammer}, H.W.}, \bibinfo{author}{{K{\"o}nig}, S.},
  \bibinfo{author}{{van Kolck}, U.}, \bibinfo{year}{2020}.
\newblock \bibinfo{title}{{Nuclear effective field theory: Status and
  perspectives}}.
\newblock \bibinfo{journal}{Reviews of Modern Physics} \bibinfo{volume}{92},
  \bibinfo{pages}{025004}.
\newblock \DOIprefix\doi{10.1103/RevModPhys.92.025004}.
\bibitem[{Henriques et~al.(1990)Henriques, Liddle and
  Moorhouse}]{henriques1990stability}
\bibinfo{author}{Henriques, A.}, \bibinfo{author}{Liddle, A.R.},
  \bibinfo{author}{Moorhouse, R.}, \bibinfo{year}{1990}.
\newblock \bibinfo{title}{Stability of boson-fermion stars}.
\newblock \bibinfo{journal}{Physics Letters B} \bibinfo{volume}{251},
  \bibinfo{pages}{511--516}.
\bibitem[{Ivanytskyi et~al.(2020)Ivanytskyi, Sagun and
  Lopes}]{ivanytskyi2020neutron}
\bibinfo{author}{Ivanytskyi, O.}, \bibinfo{author}{Sagun, V.},
  \bibinfo{author}{Lopes, I.}, \bibinfo{year}{2020}.
\newblock \bibinfo{title}{Neutron stars: New constraints on asymmetric dark
  matter}.
\newblock \bibinfo{journal}{Physical Review D} \bibinfo{volume}{102},
  \bibinfo{pages}{063028}.
\bibitem[{Kain(2020)}]{kain2020radial}
\bibinfo{author}{Kain, B.}, \bibinfo{year}{2020}.
\newblock \bibinfo{title}{Radial oscillations and stability of multiple-fluid
  compact stars}.
\newblock \bibinfo{journal}{Physical Review D} \bibinfo{volume}{102},
  \bibinfo{pages}{023001}.
\bibitem[{Kain(2021)}]{kain2021dark}
\bibinfo{author}{Kain, B.}, \bibinfo{year}{2021}.
\newblock \bibinfo{title}{Dark matter admixed neutron stars}.
\newblock \bibinfo{journal}{Physical Review D} \bibinfo{volume}{103},
  \bibinfo{pages}{043009}.
\bibitem[{{Kaplan} et~al.(2009){Kaplan}, {Luty} and {Zurek}}]{KLZ_2009}
\bibinfo{author}{{Kaplan}, D.E.}, \bibinfo{author}{{Luty}, M.A.},
  \bibinfo{author}{{Zurek}, K.M.}, \bibinfo{year}{2009}.
\newblock \bibinfo{title}{{Asymmetric dark matter}}.
\newblock \bibinfo{journal}{prd} \bibinfo{volume}{79}, \bibinfo{pages}{115016}.
\newblock \DOIprefix\doi{10.1103/PhysRevD.79.115016},
  \href{http://arxiv.org/abs/0901.4117}{{\tt arXiv:0901.4117}}.
\bibitem[{{Kievsky} et~al.(2018){Kievsky}, {Viviani}, {Logoteta}, {Bombaci} and
  {Girlanda}}]{Kievsky2018}
\bibinfo{author}{{Kievsky}, A.}, \bibinfo{author}{{Viviani}, M.},
  \bibinfo{author}{{Logoteta}, D.}, \bibinfo{author}{{Bombaci}, I.},
  \bibinfo{author}{{Girlanda}, L.}, \bibinfo{year}{2018}.
\newblock \bibinfo{title}{{Correlations imposed by the unitary limit between
  few-nucleon systems, nuclear matter, and neutron stars}}.
\newblock \bibinfo{journal}{Physical Review Letters} \bibinfo{volume}{121},
  \bibinfo{pages}{072701}.
\newblock \DOIprefix\doi{10.1103/PhysRevLett.121.072701},
  \href{http://arxiv.org/abs/1806.02636}{{\tt arXiv:1806.02636}}.
\bibitem[{Kouvaris(2008)}]{kouvaris2008wimp}
\bibinfo{author}{Kouvaris, C.}, \bibinfo{year}{2008}.
\newblock \bibinfo{title}{Wimp annihilation and cooling of neutron stars}.
\newblock \bibinfo{journal}{Physical Review D} \bibinfo{volume}{77},
  \bibinfo{pages}{023006}.
\bibitem[{Kouvaris and Nielsen(2015)}]{kouvaris2015asymmetric}
\bibinfo{author}{Kouvaris, C.}, \bibinfo{author}{Nielsen, N.G.},
  \bibinfo{year}{2015}.
\newblock \bibinfo{title}{Asymmetric dark matter stars}.
\newblock \bibinfo{journal}{Physical Review D} \bibinfo{volume}{92},
  \bibinfo{pages}{063526}.
\bibitem[{Kouvaris and Tinyakov(2010)}]{kouvaris2010can}
\bibinfo{author}{Kouvaris, C.}, \bibinfo{author}{Tinyakov, P.},
  \bibinfo{year}{2010}.
\newblock \bibinfo{title}{Can neutron stars constrain dark matter?}
\newblock \bibinfo{journal}{Physical Review D} \bibinfo{volume}{82},
  \bibinfo{pages}{063531}.
\bibitem[{Kouvaris and Tinyakov(2011)}]{kouvaris2011constraining}
\bibinfo{author}{Kouvaris, C.}, \bibinfo{author}{Tinyakov, P.},
  \bibinfo{year}{2011}.
\newblock \bibinfo{title}{Constraining asymmetric dark matter through
  observations of compact stars}.
\newblock \bibinfo{journal}{Physical Review D} \bibinfo{volume}{83},
  \bibinfo{pages}{083512}.
\bibitem[{Kouvaris et~al.(2013)}]{kouvaris2013dark}
\bibinfo{author}{Kouvaris, C.}, et~al., \bibinfo{year}{2013}.
\newblock \bibinfo{title}{The dark side of neutron stars}.
\newblock \bibinfo{journal}{Advances in High Energy Physics}
  \bibinfo{volume}{2013}.
\bibitem[{{Lattimer} et~al.(1991){Lattimer}, {Pethick}, {Prakash} and
  {Haensel}}]{durca1991}
\bibinfo{author}{{Lattimer}, J.M.}, \bibinfo{author}{{Pethick}, C.J.},
  \bibinfo{author}{{Prakash}, M.}, \bibinfo{author}{{Haensel}, P.},
  \bibinfo{year}{1991}.
\newblock \bibinfo{title}{{Direct URCA process in neutron stars}}.
\newblock \bibinfo{journal}{prl} \bibinfo{volume}{66},
  \bibinfo{pages}{2701--2704}.
\newblock \DOIprefix\doi{10.1103/PhysRevLett.66.2701}.
\bibitem[{{Lattimer} and {Prakash}(2001)}]{Lattimer-Prakash_2001}
\bibinfo{author}{{Lattimer}, J.M.}, \bibinfo{author}{{Prakash}, M.},
  \bibinfo{year}{2001}.
\newblock \bibinfo{title}{{Neutron Star Structure and the Equation of State}}.
\newblock \bibinfo{journal}{Astrophys. J.} \bibinfo{volume}{550},
  \bibinfo{pages}{426--442}.
\newblock \DOIprefix\doi{10.1086/319702},
  \href{http://arxiv.org/abs/astro-ph/0002232}{{\tt arXiv:astro-ph/0002232}}.
\bibitem[{de~Lavallaz and Fairbairn(2010)}]{de2010neutron}
\bibinfo{author}{de~Lavallaz, A.}, \bibinfo{author}{Fairbairn, M.},
  \bibinfo{year}{2010}.
\newblock \bibinfo{title}{Neutron stars as dark matter probes}.
\newblock \bibinfo{journal}{Physical Review D} \bibinfo{volume}{81},
  \bibinfo{pages}{123521}.
\bibitem[{Leung et~al.(2011)Leung, Chu and Lin}]{leung2011dark}
\bibinfo{author}{Leung, S.C.}, \bibinfo{author}{Chu, M.C.},
  \bibinfo{author}{Lin, L.M.}, \bibinfo{year}{2011}.
\newblock \bibinfo{title}{Dark-matter admixed neutron stars}.
\newblock \bibinfo{journal}{Physical Review D} \bibinfo{volume}{84},
  \bibinfo{pages}{107301}.
\bibitem[{Li et~al.(2021)Li, Cai, Xie and Zhang}]{BaoAnLi_1}
\bibinfo{author}{Li, B.A.}, \bibinfo{author}{Cai, B.J.}, \bibinfo{author}{Xie,
  W.J.}, \bibinfo{author}{Zhang, N.B.}, \bibinfo{year}{2021}.
\newblock \bibinfo{title}{Progress in constraining nuclear symmetry energy
  using neutron star observables since gw170817}.
\newblock \bibinfo{journal}{Universe} \bibinfo{volume}{7}.
\newblock \URLprefix \url{https://www.mdpi.com/2218-1997/7/6/182},
  \DOIprefix\doi{10.3390/universe7060182}.
\bibitem[{Lin and Li(2019)}]{lin2019dark}
\bibinfo{author}{Lin, H.N.}, \bibinfo{author}{Li, X.}, \bibinfo{year}{2019}.
\newblock \bibinfo{title}{The dark matter profiles in the milky way}.
\newblock \bibinfo{journal}{Monthly Notices of the Royal Astronomical Society}
  \bibinfo{volume}{487}, \bibinfo{pages}{5679--5684}.
\bibitem[{Logoteta and Bombaci(2019)}]{Logoteta:2019cyb}
\bibinfo{author}{Logoteta, D.}, \bibinfo{author}{Bombaci, I.},
  \bibinfo{year}{2019}.
\newblock \bibinfo{title}{{Constraints on Microscopic and Phenomenological
  Equations of State of Dense Matter from GW170817}}.
\newblock \bibinfo{journal}{Universe} \bibinfo{volume}{5},
  \bibinfo{pages}{204}.
\newblock \DOIprefix\doi{10.3390/universe5100204}.
\bibitem[{Logoteta et~al.(2016)Logoteta, Bombaci and Kievsky}]{LBK-2016}
\bibinfo{author}{Logoteta, D.}, \bibinfo{author}{Bombaci, I.},
  \bibinfo{author}{Kievsky, A.}, \bibinfo{year}{2016}.
\newblock \bibinfo{title}{{Nuclear matter properties from local chiral
  interactions with $\Delta$ isobar intermediate states}}.
\newblock \bibinfo{journal}{Phys. Rev. C} \bibinfo{volume}{94},
  \bibinfo{pages}{064001}.
\newblock \DOIprefix\doi{10.1103/PhysRevC.94.064001},
  \href{http://arxiv.org/abs/1609.00649}{{\tt arXiv:1609.00649}}.
\bibitem[{Logoteta et~al.(2021)Logoteta, Perego and Bombaci}]{Logoteta:2020yxf}
\bibinfo{author}{Logoteta, D.}, \bibinfo{author}{Perego, A.},
  \bibinfo{author}{Bombaci, I.}, \bibinfo{year}{2021}.
\newblock \bibinfo{title}{{Microscopic equation of state of hot nuclear matter
  for numerical relativity simulations}}.
\newblock \bibinfo{journal}{Astron. Astrophys.} \bibinfo{volume}{646},
  \bibinfo{pages}{A55}.
\newblock \DOIprefix\doi{10.1051/0004-6361/202039457},
  \href{http://arxiv.org/abs/2012.03599}{{\tt arXiv:2012.03599}}.
\bibitem[{Lopes et~al.(2019)Lopes, Lopes and Silk}]{lopes2019asteroseismology}
\bibinfo{author}{Lopes, J.}, \bibinfo{author}{Lopes, I.},
  \bibinfo{author}{Silk, J.}, \bibinfo{year}{2019}.
\newblock \bibinfo{title}{Asteroseismology of red clump stars as a probe of the
  dark matter content of the galaxy central region}.
\newblock \bibinfo{journal}{The Astrophysical Journal Letters}
  \bibinfo{volume}{880}, \bibinfo{pages}{L25}.
\bibitem[{{Machleidt} and {Entem}(2011)}]{2011PhR...503....1M}
\bibinfo{author}{{Machleidt}, R.}, \bibinfo{author}{{Entem}, D.R.},
  \bibinfo{year}{2011}.
\newblock \bibinfo{title}{{Chiral effective field theory and nuclear forces}}.
\newblock \bibinfo{journal}{Phys. Rept.} \bibinfo{volume}{503},
  \bibinfo{pages}{1--75}.
\newblock \DOIprefix\doi{10.1016/j.physrep.2011.02.001},
  \href{http://arxiv.org/abs/1105.2919}{{\tt arXiv:1105.2919}}.
\bibitem[{Mariani et~al.(2024)Mariani, Albertus, Alessandroni, Orsaria,
  P{\'e}rez-Garc{\'\i}a and Ranea-Sandoval}]{mariani2024constraining}
\bibinfo{author}{Mariani, M.}, \bibinfo{author}{Albertus, C.},
  \bibinfo{author}{Alessandroni, M.d.R.}, \bibinfo{author}{Orsaria, M.G.},
  \bibinfo{author}{P{\'e}rez-Garc{\'\i}a, M.{\'A}.},
  \bibinfo{author}{Ranea-Sandoval, I.F.}, \bibinfo{year}{2024}.
\newblock \bibinfo{title}{Constraining self-interacting fermionic dark matter
  in admixed neutron stars using multimessenger astronomy}.
\newblock \bibinfo{journal}{Monthly Notices of the Royal Astronomical Society}
  \bibinfo{volume}{527}, \bibinfo{pages}{6795--6806}.
\bibitem[{{Merritt} et~al.(2006){Merritt}, {Graham}, {Moore}, {Diemand} and
  {Terzi{\'c}}}]{merritt2006}
\bibinfo{author}{{Merritt}, D.}, \bibinfo{author}{{Graham}, A.W.},
  \bibinfo{author}{{Moore}, B.}, \bibinfo{author}{{Diemand}, J.},
  \bibinfo{author}{{Terzi{\'c}}, B.}, \bibinfo{year}{2006}.
\newblock \bibinfo{title}{{Empirical Models for Dark Matter Halos. I.
  Nonparametric Construction of Density Profiles and Comparison with Parametric
  Models}}.
\newblock \bibinfo{journal}{Astronomical Journal} \bibinfo{volume}{132},
  \bibinfo{pages}{2685--2700}.
\newblock \DOIprefix\doi{10.1086/508988},
  \href{http://arxiv.org/abs/astro-ph/0509417}{{\tt arXiv:astro-ph/0509417}}.
\bibitem[{Miller et~al.(2019)Miller, Lamb, Dittmann, Bogdanov, Arzoumanian,
  Gendreau, Guillot, Harding, Ho, Lattimer et~al.}]{miller2019psr}
\bibinfo{author}{Miller, M.}, \bibinfo{author}{Lamb, F.K.},
  \bibinfo{author}{Dittmann, A.}, \bibinfo{author}{Bogdanov, S.},
  \bibinfo{author}{Arzoumanian, Z.}, \bibinfo{author}{Gendreau, K.C.},
  \bibinfo{author}{Guillot, S.}, \bibinfo{author}{Harding, A.},
  \bibinfo{author}{Ho, W.}, \bibinfo{author}{Lattimer, J.}, et~al.,
  \bibinfo{year}{2019}.
\newblock \bibinfo{title}{Psr j0030+ 0451 mass and radius from nicer data and
  implications for the properties of neutron star matter}.
\newblock \bibinfo{journal}{The Astrophysical Journal Letters}
  \bibinfo{volume}{887}, \bibinfo{pages}{L24}.
\bibitem[{Miller et~al.(2021)}]{Miller:2021qha}
\bibinfo{author}{Miller, M.C.}, et~al., \bibinfo{year}{2021}.
\newblock \bibinfo{title}{{The Radius of PSR J0740+6620 from NICER and
  XMM-Newton Data}}.
\newblock \bibinfo{journal}{Astrophys. J. Lett.} \bibinfo{volume}{918},
  \bibinfo{pages}{L28}.
\newblock \DOIprefix\doi{10.3847/2041-8213/ac089b},
  \href{http://arxiv.org/abs/2105.06979}{{\tt arXiv:2105.06979}}.
\bibitem[{Narain et~al.(2006)Narain, Schaffner-Bielich and
  Mishustin}]{narain2006compact}
\bibinfo{author}{Narain, G.}, \bibinfo{author}{Schaffner-Bielich, J.},
  \bibinfo{author}{Mishustin, I.N.}, \bibinfo{year}{2006}.
\newblock \bibinfo{title}{Compact stars made of fermionic dark matter}.
\newblock \bibinfo{journal}{Physical Review D} \bibinfo{volume}{74},
  \bibinfo{pages}{063003}.
\bibitem[{{Negele} and {Vautherin}(1973)}]{NV73}
\bibinfo{author}{{Negele}, J.W.}, \bibinfo{author}{{Vautherin}, D.},
  \bibinfo{year}{1973}.
\newblock \bibinfo{title}{{Neutron star matter at sub-nuclear densities}}.
\newblock \bibinfo{journal}{Nucl. Phys. A} \bibinfo{volume}{207},
  \bibinfo{pages}{298--320}.
\newblock \DOIprefix\doi{10.1016/0375-9474(73)90349-7}.
\bibitem[{Nelson et~al.(2019)Nelson, Reddy and Zhou}]{nelson2019dark}
\bibinfo{author}{Nelson, A.E.}, \bibinfo{author}{Reddy, S.},
  \bibinfo{author}{Zhou, D.}, \bibinfo{year}{2019}.
\newblock \bibinfo{title}{Dark halos around neutron stars and gravitational
  waves}.
\newblock \bibinfo{journal}{Journal of Cosmology and Astroparticle Physics}
  \bibinfo{volume}{2019}, \bibinfo{pages}{012}.
\bibitem[{Nesti et~al.(2023)Nesti, Salucci and Turini}]{nesti2023quest}
\bibinfo{author}{Nesti, F.}, \bibinfo{author}{Salucci, P.},
  \bibinfo{author}{Turini, N.}, \bibinfo{year}{2023}.
\newblock \bibinfo{title}{The quest for the nature of the dark matter: The need
  of a new paradigm}.
\newblock \bibinfo{journal}{Astronomy} \bibinfo{volume}{2},
  \bibinfo{pages}{90--104}.
\bibitem[{{Page} et~al.(2004){Page}, {Lattimer}, {Prakash} and
  {Steiner}}]{Page2004}
\bibinfo{author}{{Page}, D.}, \bibinfo{author}{{Lattimer}, J.M.},
  \bibinfo{author}{{Prakash}, M.}, \bibinfo{author}{{Steiner}, A.W.},
  \bibinfo{year}{2004}.
\newblock \bibinfo{title}{{Minimal Cooling of Neutron Stars: A New Paradigm}}.
\newblock \bibinfo{journal}{The Astrophysical Journal Supplement Series}
  \bibinfo{volume}{155}, \bibinfo{pages}{623--650}.
\newblock \DOIprefix\doi{10.1086/424844},
  \href{http://arxiv.org/abs/astro-ph/0403657}{{\tt arXiv:astro-ph/0403657}}.
\bibitem[{Pato et~al.(2015)Pato, Iocco and Bertone}]{pato2015dynamical}
\bibinfo{author}{Pato, M.}, \bibinfo{author}{Iocco, F.},
  \bibinfo{author}{Bertone, G.}, \bibinfo{year}{2015}.
\newblock \bibinfo{title}{Dynamical constraints on the dark matter distribution
  in the milky way}.
\newblock \bibinfo{journal}{Journal of Cosmology and Astroparticle Physics}
  \bibinfo{volume}{2015}, \bibinfo{pages}{001}.
\bibitem[{{Peebles}(1982)}]{Peebels1982}
\bibinfo{author}{{Peebles}, P.J.E.}, \bibinfo{year}{1982}.
\newblock \bibinfo{title}{{Large-scale background temperature and mass
  fluctuations due to scale-invariant primeval perturbations}}.
\newblock \bibinfo{journal}{The Astrophysical Journal Letters}
  \bibinfo{volume}{263}, \bibinfo{pages}{L1--L5}.
\newblock \DOIprefix\doi{10.1086/183911}.
\bibitem[{Prakash et~al.(2021)Prakash, Radice, Logoteta, Perego, Nedora,
  Bombaci, Kashyap, Bernuzzi and Endrizzi}]{A.Prakash_2021}
\bibinfo{author}{Prakash, A.}, \bibinfo{author}{Radice, D.},
  \bibinfo{author}{Logoteta, D.}, \bibinfo{author}{Perego, A.},
  \bibinfo{author}{Nedora, V.}, \bibinfo{author}{Bombaci, I.},
  \bibinfo{author}{Kashyap, R.}, \bibinfo{author}{Bernuzzi, S.},
  \bibinfo{author}{Endrizzi, A.}, \bibinfo{year}{2021}.
\newblock \bibinfo{title}{Signatures of deconfined quark phases in binary
  neutron star mergers}.
\newblock \bibinfo{journal}{PHYSICAL REVIEW D} \bibinfo{volume}{104}.
\newblock \DOIprefix\doi{10.1103/PhysRevD.104.083029}.
\bibitem[{Raaijmakers et~al.(2021)Raaijmakers, Greif, Hebeler, Hinderer,
  Nissanke, Schwenk, Riley, Watts, Lattimer and
  Ho}]{raaijmakers2021constraints}
\bibinfo{author}{Raaijmakers, G.}, \bibinfo{author}{Greif, S.},
  \bibinfo{author}{Hebeler, K.}, \bibinfo{author}{Hinderer, T.},
  \bibinfo{author}{Nissanke, a.}, \bibinfo{author}{Schwenk, A.},
  \bibinfo{author}{Riley, T.}, \bibinfo{author}{Watts, A.},
  \bibinfo{author}{Lattimer, J.}, \bibinfo{author}{Ho, W.},
  \bibinfo{year}{2021}.
\newblock \bibinfo{title}{Constraints on the dense matter equation of state and
  neutron star properties from nicer’s mass--radius estimate of psr j0740+
  6620 and multimessenger observations}.
\newblock \bibinfo{journal}{The Astrophysical Journal Letters}
  \bibinfo{volume}{918}, \bibinfo{pages}{L29}.
\bibitem[{Reed et~al.(2021)Reed, Fattoyev, Horowitz and
  Piekarewicz}]{PhysRevLett.126.172503}
\bibinfo{author}{Reed, B.T.}, \bibinfo{author}{Fattoyev, F.J.},
  \bibinfo{author}{Horowitz, C.J.}, \bibinfo{author}{Piekarewicz, J.},
  \bibinfo{year}{2021}.
\newblock \bibinfo{title}{Implications of prex-2 on the equation of state of
  neutron-rich matter}.
\newblock \bibinfo{journal}{Phys. Rev. Lett.} \bibinfo{volume}{126},
  \bibinfo{pages}{172503}.
\newblock \URLprefix
  \url{https://link.aps.org/doi/10.1103/PhysRevLett.126.172503},
  \DOIprefix\doi{10.1103/PhysRevLett.126.172503}.
\bibitem[{Scott et~al.(2009)Scott, Fairbairn and Edsj{\"o}}]{scott2009dark}
\bibinfo{author}{Scott, P.}, \bibinfo{author}{Fairbairn, M.},
  \bibinfo{author}{Edsj{\"o}, J.}, \bibinfo{year}{2009}.
\newblock \bibinfo{title}{Dark stars at the galactic centre--the main
  sequence}.
\newblock \bibinfo{journal}{Monthly Notices of the Royal Astronomical Society}
  \bibinfo{volume}{394}, \bibinfo{pages}{82--104}.
\bibitem[{{Shakeri} and {Karkevandi}(2024)}]{Shakeri-Karkevandi2024}
\bibinfo{author}{{Shakeri}, S.}, \bibinfo{author}{{Karkevandi}, D.R.},
  \bibinfo{year}{2024}.
\newblock \bibinfo{title}{{Bosonic dark matter in light of the NICER precise
  mass-radius measurements}}.
\newblock \bibinfo{journal}{prd} \bibinfo{volume}{109},
  \bibinfo{pages}{043029}.
\newblock \DOIprefix\doi{10.1103/PhysRevD.109.043029},
  \href{http://arxiv.org/abs/2210.17308}{{\tt arXiv:2210.17308}}.
\bibitem[{{Shapiro} and {Teukolsky}(1983)}]{Shapiro-Teukolsky_1983}
\bibinfo{author}{{Shapiro}, S.L.}, \bibinfo{author}{{Teukolsky}, S.A.},
  \bibinfo{year}{1983}.
\newblock \bibinfo{title}{{Black holes, white dwarfs and neutron stars. The
  physics of compact objects}}.
\newblock \DOIprefix\doi{10.1002/9783527617661}.
\bibitem[{Sofue(2013)}]{sofue2013rotation}
\bibinfo{author}{Sofue, Y.}, \bibinfo{year}{2013}.
\newblock \bibinfo{title}{Rotation curve and mass distribution in the galactic
  center—from black hole to entire galaxy—}.
\newblock \bibinfo{journal}{Publications of the Astronomical Society of Japan}
  \bibinfo{volume}{65}, \bibinfo{pages}{118}.
\bibitem[{{Steiner}(2006)}]{Steiner_2006}
\bibinfo{author}{{Steiner}, A.W.}, \bibinfo{year}{2006}.
\newblock \bibinfo{title}{{High-density symmetry energy and direct Urca
  process}}.
\newblock \bibinfo{journal}{Phys. Rev. C} \bibinfo{volume}{74},
  \bibinfo{pages}{045808}.
\newblock \DOIprefix\doi{10.1103/PhysRevC.74.045808},
  \href{http://arxiv.org/abs/nucl-th/0607040}{{\tt arXiv:nucl-th/0607040}}.
\bibitem[{{Tews} et~al.(2017){Tews}, {Lattimer}, {Ohnishi} and
  {Kolomeitsev}}]{Tews2017}
\bibinfo{author}{{Tews}, I.}, \bibinfo{author}{{Lattimer}, J.M.},
  \bibinfo{author}{{Ohnishi}, A.}, \bibinfo{author}{{Kolomeitsev}, E.E.},
  \bibinfo{year}{2017}.
\newblock \bibinfo{title}{{Symmetry Parameter Constraints from a Lower Bound on
  Neutron-matter Energy}}.
\newblock \bibinfo{journal}{Astrophysical Journal} \bibinfo{volume}{848},
  \bibinfo{pages}{105}.
\newblock \DOIprefix\doi{10.3847/1538-4357/aa8db9},
  \href{http://arxiv.org/abs/1611.07133}{{\tt arXiv:1611.07133}}.
\bibitem[{Thakur et~al.(2024)Thakur, Malik, Das, Jha, Sharma and
  Provid{\^e}ncia}]{thakur2024feasibility}
\bibinfo{author}{Thakur, P.}, \bibinfo{author}{Malik, T.},
  \bibinfo{author}{Das, A.}, \bibinfo{author}{Jha, T.},
  \bibinfo{author}{Sharma, B.}, \bibinfo{author}{Provid{\^e}ncia, C.},
  \bibinfo{year}{2024}.
\newblock \bibinfo{title}{Feasibility of dark matter admixed neutron star based
  on recent observational constraints}.
\newblock \bibinfo{journal}{arXiv preprint arXiv:2408.03780} .
\bibitem[{Tolos and Schaffner-Bielich(2015)}]{tolos2015dark}
\bibinfo{author}{Tolos, L.}, \bibinfo{author}{Schaffner-Bielich, J.},
  \bibinfo{year}{2015}.
\newblock \bibinfo{title}{Dark compact planets}.
\newblock \bibinfo{journal}{Physical Review D} \bibinfo{volume}{92},
  \bibinfo{pages}{123002}.
\bibitem[{Valdez-Alvarado et~al.(2013)Valdez-Alvarado, Palenzuela, Alic and
  Urena-L{\'o}pez}]{valdez2013dynamical}
\bibinfo{author}{Valdez-Alvarado, S.}, \bibinfo{author}{Palenzuela, C.},
  \bibinfo{author}{Alic, D.}, \bibinfo{author}{Urena-L{\'o}pez, L.A.},
  \bibinfo{year}{2013}.
\newblock \bibinfo{title}{Dynamical evolution of fermion-boson stars}.
\newblock \bibinfo{journal}{Physical Review D} \bibinfo{volume}{87},
  \bibinfo{pages}{084040}.
\bibitem[{Zhang and Li(2021)}]{BaoAnLi_2}
\bibinfo{author}{Zhang, N.B.}, \bibinfo{author}{Li, B.A.},
  \bibinfo{year}{2021}.
\newblock \bibinfo{title}{Impact of nicer’s radius measurement of psr
  j0740+6620 on nuclear symmetry energy at suprasaturation densities}.
\newblock \bibinfo{journal}{The Astrophysical Journal} \bibinfo{volume}{921},
  \bibinfo{pages}{111}.
\newblock \URLprefix \url{https://dx.doi.org/10.3847/1538-4357/ac1e8c},
  \DOIprefix\doi{10.3847/1538-4357/ac1e8c}.
\bibitem[{{Zurek}(2014)}]{Zurek_2014}
\bibinfo{author}{{Zurek}, K.M.}, \bibinfo{year}{2014}.
\newblock \bibinfo{title}{{Asymmetric Dark Matter: Theories, signatures, and
  constraints}}.
\newblock \bibinfo{journal}{Phys. Rep.} \bibinfo{volume}{537},
  \bibinfo{pages}{91--121}.
\newblock \DOIprefix\doi{10.1016/j.physrep.2013.12.001},
  \href{http://arxiv.org/abs/1308.0338}{{\tt arXiv:1308.0338}}.

\end{thebibliography}
\end{document}